\documentclass[a4paper,onecolumn,11pt,accepted=2025-04-29]{quantumarticle}
\pdfoutput=1
\usepackage[utf8]{inputenc}
\usepackage[english]{babel}
\usepackage[T1]{fontenc}
\usepackage{amsmath}
\usepackage{hyperref}

\usepackage{tikz}
\usepackage{lipsum}

\begin{document}

\title{Spontaneous symmetry emergence in a Hermitian system of coupled oscillators without symmetry}

\author{T. T. Sergeev}
\affiliation{Dukhov Research Institute of Automatics, 127055,  22 Sushchevskaya, Moscow Russia}
\affiliation{Moscow Institute of Physics and Technology, 141700, 9 Institutskiy pereulok, Dolgoprudny Moscow region, Russia}
\affiliation{Institute for Theoretical and Applied Electromagnetics, 125412, 13 Izhorskaya, Moscow Russia}

\author{E. S. Andrianov}
\affiliation{Dukhov Research Institute of Automatics, 127055, 22 Sushchevskaya, Moscow Russia}
\affiliation{Moscow Institute of Physics and Technology, 141700, 9 Institutskiy pereulok, Dolgoprudny Moscow region, Russia}
\affiliation{Institute for Theoretical and Applied Electromagnetics, 125412, 13 Izhorskaya, Moscow Russia}

\author{A. A. Zyablovsky}
\email{zyablovskiy@mail.ru}
\affiliation{Dukhov Research Institute of Automatics, 127055, 22 Sushchevskaya, Moscow Russia}
\affiliation{Moscow Institute of Physics and Technology, 141700, 9 Institutskiy pereulok, Dolgoprudny Moscow region, Russia}
\affiliation{Institute for Theoretical and Applied Electromagnetics, 125412, 13 Izhorskaya, Moscow Russia}

\maketitle

\begin{abstract}
  Spontaneous symmetry breaking in systems with symmetry is a cornerstone phenomenon accompanying second-order phase transitions. Here, we predict the opposite phenomenon, namely, spontaneous symmetry emergence in a system that lacks symmetry. In the example of two coupled oscillators interacting non-symmetrically with a set of oscillators whose frequencies uniformly fill a finite frequency range, we demonstrate that the system state can acquire symmetry that is not inherent in the system Hamiltonian. The emergence of symmetry is manifested as a change in the system dynamics, which can be interpreted as a phase transition in a Hermitian system that lacks symmetry.
\end{abstract}

\section{Introduction}
Spontaneous symmetry breaking is a key concept in physics~\cite{ref1,ref2,ref43N,ref44N} and is used for the description of second-order phase transitions~\cite{ref1,ref2,ref43N,ref44N}. Conventionally, phase transitions are studied in closed systems described by Hermitian Hamiltonians~\cite{ref3}. It is known~\cite{ref3} that if the linear Hermitian operator commutes with any other linear Hermitian operator, then one can find a system of eigenstates that is common to both operators. Therefore, if the system Hamiltonian has symmetry, i.e., commutes with the symmetry operator, then there is a basis of eigenstates that are also eigenstates of the symmetry operator~\cite{ref3}. A phase transition occurs when the symmetry of the instantaneous state of the system becomes lower than the Hamiltonian symmetry, which is possible when the ground state of the system is degenerate~\cite{ref3}. In this case, the spontaneous symmetry breaking means a decrease in the symmetry of the instantaneous state of the system~\cite{ref1}, though the symmetry of the eigenstates coincides with the Hamiltonian symmetry.

In recent decades, it has been shown that in non-Hermitian systems, it is possible that the symmetry of the eigenstates is lower than the symmetry of the system Hamiltonian~\cite{ref4,ref5,ref6}. As a result, one can observe another type of spontaneous symmetry breaking~\cite{ref6,ref7,ref8,ref9,ref10}, which manifests as the decrease of the symmetry of eigenstates with simultaneous conservation of the Hamiltonian symmetry. This type of spontaneous symmetry breaking is associated with non-Hermitian phase transitions~\cite{ref11,ref12,ref13,ref14,ref15,ref16,ref17,ref18}. PT-symmetrical systems are one of the most famous examples of systems with this type of spontaneous symmetry breaking~\cite{ref13,ref14}. In these systems, the change in the system parameters leads to passing through an exceptional point (EP)~\cite{ref13,ref14}, at which the eigenstates cease to be PT-symmetrical, whereas the Hamiltonian is still PT-symmetrical. In addition to PT-symmetrical systems, non-Hermitian phase transitions can also occur in strongly coupled cavity-atom systems~\cite{ref13,ref19}, polariton~\cite{ref20,ref21}, optomechanical~\cite{ref23,ref24,ref26,ref27,ref28,ref29}, and laser~\cite{ref25,ref31,ref33,ref34,ref35} systems. The systems with the phase transitions at the EP find a number of the applications~\cite{ref22,ref30,ref32,liertzer2012,longhi2009,suchkov2016}.

However, a consistent description of the phase transitions at the EPs encounters difficulty. Indeed, any non-Hermitian system is actually subsystem of a larger Hermitian system, consisting of the considered non-Hermitian system and its environment. The transition from a Hermitian description to a non-Hermitian one is carried out by elimination of the environment's degrees of freedom, in particular, in the Born-Markovian approximation. The presence of the symmetry in the resulting non-Hermitian system does not guarantee that the original Hermitian system has the same symmetry. That is, the symmetry of a non-Hermitian system and its eigenstates may be a consequence of the inaccuracy of non-Hermitian description. Consequently, the spontaneous symmetry breaking of eigenstates can also only be a consequence of the inaccuracy of the non-Hermitian consideration.

An Hermitian system that describe PT-symmetry breaking after elimination reservoir's degrees of freedom is a system consisting of two coupled oscillators and a set of oscillators interacting with one of the coupled oscillators. Such a Hermitian system is not symmetric with respect to the permutation of the first and second oscillators. However, if we exclude from consideration the degrees of freedom of the oscillators in the set then the resulting non-Hermitian system will be quasi-PT-symmetrical \cite{ref8}. The eigenstates of such a non-Hermitian system are PT-symmetrical when the coupling strength between the oscillators is greater than a threshold value. Below the threshold values, the eigenstates are not PT-symmetrical and therefore a spontaneous breaking of PT-symmetry of the eigenstates occurs at the threshold value of coupling strength. However, since symmetry is not inherent in the initial system, it is unclear whether spontaneous symmetry breaking is associated with any change in the Hermitian system.

Here, we demonstrate that in the Hermitian system that is not symmetrical with respect to permutation of its elements the phenomenon opposite to spontaneous symmetry breaking, namely, spontaneous symmetry emergence takes place. On the example of two harmonic oscillators one of which interacts with the set of large number of harmonic oscillators, we show that an increase in the coupling strength between the oscillators leads to the emergence of symmetry with respect to the permutation of the first and second oscillators in the dynamics of the Hermitian system. We demonstrate that there is a critical coupling strength above which the average absolute amplitudes of the first and second oscillators become equal to each other. This is associated with the appearance of a sharp threshold and is a manifestation of a new phenomenon that can be called spontaneous symmetry emergence. Thus, we predict the existence of a spontaneous symmetry emergence that takes place in the system, which does not possess symmetry.

The critical coupling strength coincides with the one, at which the splitting in the spectrum of the quasi-PT-symmetrical system appears. Based on them, we can assume that spontaneous symmetry emergence in the Hermitian systems can be an initial cause of the changing of symmetry in the non-Hermitian systems. We believe that the spontaneous symmetry emergence serves as a source for phase transitions in a Hermitian systems that lacks symmetry.

\section{System under consideration}
\begin{figure}[htbp]
\centering\includegraphics[width=0.7\linewidth]{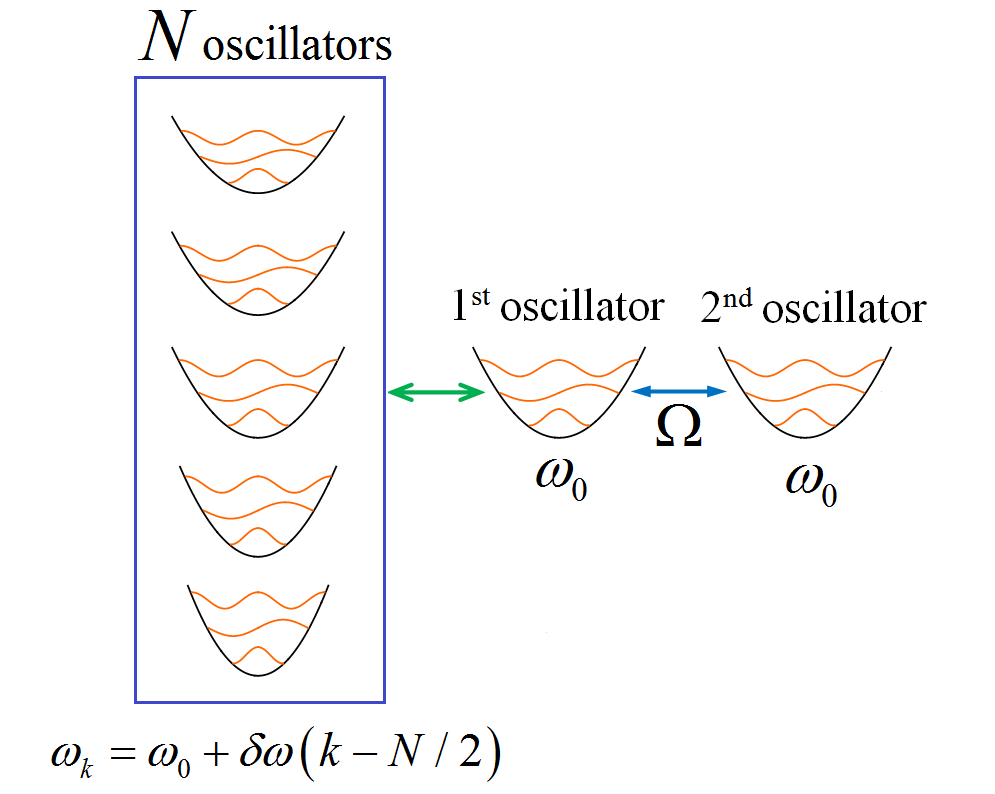}
\caption{The scheme of the system under consideration.}
\label{fig1}
\end{figure}

We consider a Hermitian system consisting of two coupled oscillators and a set of $N$  oscillators interacting with the first oscillator [Figure~\ref{fig1}]. The two main oscillators have equal frequencies, ${\omega _0}$. The frequencies of oscillators in the set are ${\omega _k} = {\omega _0} + \delta \omega \,\left( {k - N/2} \right)$, where $k$ is a number of the oscillators and $\delta \omega  <  < {\omega _0}$ is a distance between the frequencies of the oscillators. In the experiment, this system can be realized, for example, on the basis of an optical structure of two coupled single-mode cavities, one of which interacts with a waveguide of finite length. Due to the finite length of the waveguide, its modes have a discrete set of frequencies, the specific values of which are determined by the frequency dispersion and geometry of the waveguide.

We use the following Hamiltonian for the system~\cite{ref37}:

\begin{equation}
\begin{array}{l}
\hat H = {\omega _0}\hat a_1^\dag {{\hat a}_1} + {\omega _0}\hat a_2^\dag {{\hat a}_2} + \Omega (\hat a_1^\dag {{\hat a}_2} + \hat a_2^\dag {{\hat a}_1}) + \\
\sum\limits_{k = 1}^N {{\omega _k}\hat b_k^\dag {{\hat b}_k}}  + \sum\limits_{k = 1}^N {g(\hat b_k^\dag {{\hat a}_1} + \hat a_1^\dag {{\hat b}_k})} 
\end{array}
\label{eq:refname1}
\end{equation}
where ${\hat a_{1,2}}$ and $\hat a_{1,2}^\dag$ are the annihilation and creation operators of the first and second oscillators, obeying the boson commutation relations $\left[ {{{\hat a}_i},\hat a_j^\dag } \right] = {\delta _{ij}}\hat 1$,  $i,j = 1,2$~\cite{ref37}. ${\hat b_k}$ and $\hat b_k^\dag$ are the annihilation and creation operators of the oscillators in the set with the frequencies ${\omega _k}$. $\Omega$ is a coupling strength between the first and second oscillators; $g$ is a coupling strength between the first oscillator and each of the oscillators in the set. Note that we use the rotating-wave approximation to describe the interaction between oscillators~\cite{ref37}.

There are several approaches to studying the dynamics of the system~(\ref{eq:refname1}). One approach was proposed in the work of Caldeira and Leggett~\cite{ref45N} and is based on the study of the total density matrix using the Feynman-Vernon functional integral~\cite{ref45N,ref46N,ref41}. Another approach is to use Heisenberg equations for the operators \cite{ref38,ref39}. In our case, the last approach is more suitable because it enables us to consider the oscillators' amplitudes without additional approximations. Using the Heisenberg equation for operators~\cite{ref38,ref39}, we obtain a closed system of equations for operators ${\hat a_{1,2}}$, ${\hat b_k}$. Then moving from the operators to their averages, we obtain the linear system of equations (see Appendix A for more details)

\begin{equation}
\frac{{d{a_1}}}{{dt}} =  - i{\omega _0}{a_1} - i\,\Omega {a_2} - i\sum\limits_{k = 1}^N {g{b_k}}
\label{eq:refname2}
\end{equation}

\begin{equation}
\frac{{d{a_2}}}{{dt}} =  - i{\omega _0}{a_2} - i\,\Omega {a_1}
\label{eq:refname3}
\end{equation}

\begin{equation}
\frac{{d{b_k}}}{{dt}} =  - i{\omega _k}{b_k} - ig{a_1}
\label{eq:refname4}
\end{equation}
where ${a_1} = \left\langle {{{\hat a}_1}} \right\rangle$, ${a_2} = \left\langle {{{\hat a}_2}} \right\rangle$ and ${b_k} = \left\langle {{{\hat b}_k}} \right\rangle$.

\section{Non-Hermitian description of the system}

The interaction of the first oscillator with the set of oscillators leads to an energy exchange between them. When the step between the frequencies of oscillators in the set tends to zero ($\delta \omega /{\omega _0} \to 0$) and the number of the oscillators in the set tends to infinity ($N \to \infty$ and $N \delta \omega = const$), this energy exchange transforms into energy dissipation in the first oscillator. In this case, the degrees of freedom of the oscillators in the set can be eliminated from consideration with the aid of the Born-Markovian approximation~\cite{ref38,ref39}. In this approach, the system dynamics are described by the stochastic non-Hermitian equation (see Appendices B and C):

\begin{equation}
\frac{d}{{dt}}\left( {\begin{array}{*{20}{c}}
{{a_1}}\\
{{a_2}}
\end{array}} \right) = \left( {\begin{array}{*{20}{c}}
{ - i{\omega _0} - \gamma }&{ - i\Omega }\\
{ - i\Omega }&{ - i{\omega _0}}
\end{array}} \right)\left( {\begin{array}{*{20}{c}}
{{a_1}}\\
{{a_2}}
\end{array}} \right) + \left( {\begin{array}{*{20}{c}}
{{\xi}}\\
{{0}}
\end{array}} \right)
\label{eq:refname5}
\end{equation}
where $\gamma  = \pi {g^2}/\delta \omega$ is an effective decay rate and $\xi$ is a noise term, the correlation function of which is determined by the fluctuation-dissipation theorem~\cite{ref37,ref38,ref39}. 

To describe the system behavior is often used the non-Hermitian equations~(\ref{eq:refname5}) without noise term. The eigenvalues of the equations~(\ref{eq:refname5}) without noise term are ${\lambda _ \pm } =  - \frac{\gamma }{2} - i \omega_0 \pm \frac{1}{2}\sqrt {{\gamma ^2} - 4\,{\Omega ^2}}$ and the respective eigenstates have the form:

\begin{equation}
{{\bf{e}}_ \pm } = {\left\{ {i\left( { - \gamma  \pm \sqrt {{\gamma ^2} - 4\,{\Omega ^2}} } \right)/2\Omega \,,\,\,\,1} \right\}^T}
\label{eq:refname1N}
\end{equation}
In this non-Hermitian system, there is an exceptional point (EP) when $\Omega  = {\Omega _{EP}} = \gamma /2$. At the EP, the eigenstates of non-Hermitian system become linearly dependent, and their eigenvalues are equal to each other~\cite{ref13}. When $\Omega  < {\Omega _{EP}}$, the absolute values of the first and second components of each eigenstates are not equal to each other ($\left| {{{\left( {{{\bf{e}}_ \pm }} \right)}_1}} \right| \ne \left| {{{\left( {{{\bf{e}}_ \pm }} \right)}_2}} \right|$) and the eigenstates~(\ref{eq:refname1N}) are not PT-symmetrical. In this case, the imaginary parts of the eigenvalues, which determine the oscillation frequencies of the eigenstates, are equal to each other. When $\Omega  > {\Omega _{EP}}$, the absolute values of the first and second components of the eigenstates become equal to each other ($\left| {{{\left( {{{\bf{e}}_ \pm }} \right)}_1}} \right| = \left| {{{\left( {{{\bf{e}}_ \pm }} \right)}_2}} \right|$) and the eigenstates~(\ref{eq:refname1N}) are PT-symmetrical \cite{ref14}. In this case, the imaginary parts of the eigenvalues differ from each other. This indicates splitting in the system spectrum at the EP. The passing through the EP, occurring at the increase of coupling strength, is often associated with the spontaneous symmetry breaking in the eigenstates~\cite{ref8,ref13}.

\begin{figure}[htbp]
\centering\includegraphics[width=0.7\linewidth]{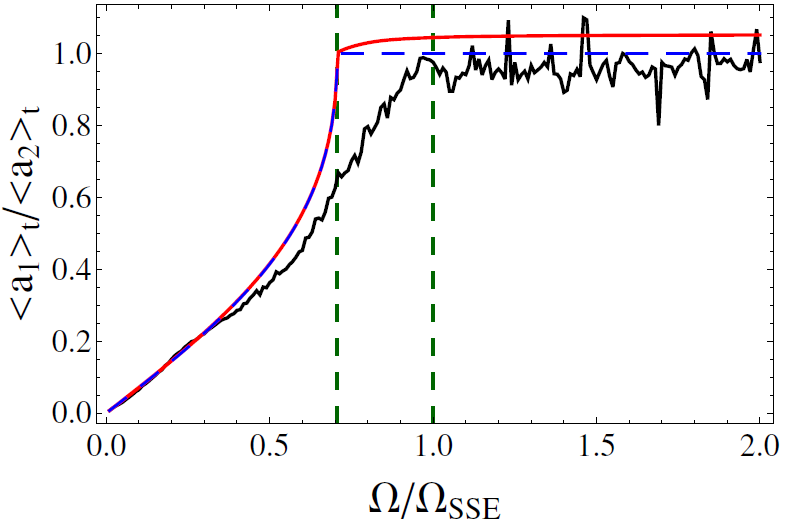}
\caption{The dependence of ${\left\langle {\left| {{a_1}} \right|} \right\rangle _t}/{\left\langle {\left| {{a_2}} \right|} \right\rangle _t}$ on the coupling strength $\Omega$ when the initial state has the form ${a_1}\left( {t = 0} \right) = {\left( {{{\bf{e}}_ + }} \right)_1}$, ${a_2}\left( {t = 0} \right) = {\left( {{{\bf{e}}_ + }} \right)_2}$  (see Eq.~(\ref{eq:refname1N})), and ${b_k}\left( {t = 0} \right) = 0$. The dashed blue line is calculated by the Eq.~(\ref{eq:refname5}) and $t_{max} \to \infty$. The solid lines are calculated by the Eqns.~(\ref{eq:refname2})-(\ref{eq:refname4}) when ${\left\langle {\left| {{a_1}} \right|} \right\rangle _t}/{\left\langle {\left| {{a_2}} \right|} \right\rangle _t}$ is averaged over the time interval $0 \leq t \leq t_{max} = 0.5{T_R}$ (the red line) and $0 \leq t \leq t_{max} = 10{T_R}$ (the black line). The dashed vertical lines show $\Omega={\Omega _{EP}} = \gamma /2 $ and $\Omega={\Omega _{SSE}} = \gamma /\sqrt 2 $. Here $\Omega_{SSE}$ is a coupling strength, at which splitting in the cmponents of the eigenstates of the Hermtian system~(\ref{eq:refname2})-(\ref{eq:refname4}) begins to appear. $T_R = 2 \pi / \delta \omega$ is the return time. $\gamma  = \pi {g^2}/\delta \omega \approx 0.014\omega_0$, $\delta\omega = 0.001\omega_0$, $g=\sqrt{2}\cdot0.0015\omega_0$ (for $N=200$) and we consider $\omega_0=1$.}
\label{fig3}
\end{figure}

The eigenstates can be symmetrical because the non-Hermitian system described by Eq.~(\ref{eq:refname5}) is quasi-PT-symmetrical \cite{ref8,ref14}. It is seen that adding to the matrix on the right side of~(\ref{eq:refname5}) the identity matrix multiplied by $\gamma /2$, we obtain the following matrix:

\begin{equation}
M = \left( {\begin{array}{*{20}{c}}
  { - i{\omega _0} - \gamma /2}&{ - i\Omega } \\ 
  { - i\Omega }&{ - i{\omega _0} + \gamma /2} 
\end{array}} \right)
\label{eq:refname1NN}
\end{equation}
which is PT-symmetrical \cite{ref14}. Due to the fact that $M$ is a non-Hermitian matrix, it is possible that the symmetry of the eigenstates is lower than the symmetry of the system matrix \cite{ref5,ref6}. As a result, one can observe spontaneous symmetry breaking in the eigenvectors \cite{ref7,ref8}, when the eigenvectors are PT-symmetrical at $\Omega  > {\Omega _{EP}}$ and non-PT-symmetrical at $\Omega  < {\Omega _{EP}}$. Adding the identity matrix does not change the eigenvectors of any matrix. Therefore, in the non-Hermitian system described by Eq.~(\ref{eq:refname5}) the spontaneous symmetry breaking in the eigenstates is also observed. The transitions related with the spontaneous PT-symmetry breaking can occur in quantum \cite{ref4,bender1999}, optical \cite{ref13,ref14}, magnonic \cite{magnonReview,ref18,wang2022magnon,ref21,ref22}, Bose–Einstein condensate \cite{jack1996coherent,trimborn2008mean,graefe,konotop} systems. These transitions manifest themselves in the system's dynamics. In particular, if the initial state of the system coincides with one of the eigenstates then the ratio of the absolute values of the amplitudes of the first and second oscillators averaged over time, ${\left\langle {\left| {{a_1}} \right|} \right\rangle _t}/{\left\langle {\left| {{a_2}} \right|} \right\rangle _t}$, becomes equal to $1$ at $\Omega  > {\Omega _{EP}}$ [see the dashed blue line in Figure~\ref{fig3}], where ${\left\langle {\left| {{a_{1,2}}} \right|} \right\rangle _t} = \frac{1}{t_{max}}\int_{0}^{t_{max}} {\left| {{a_{1,2}}} \right|} \,dt$ and $t_{max} \to \infty$.

\begin{figure*}[htbp]
\centering\includegraphics[width=\linewidth]{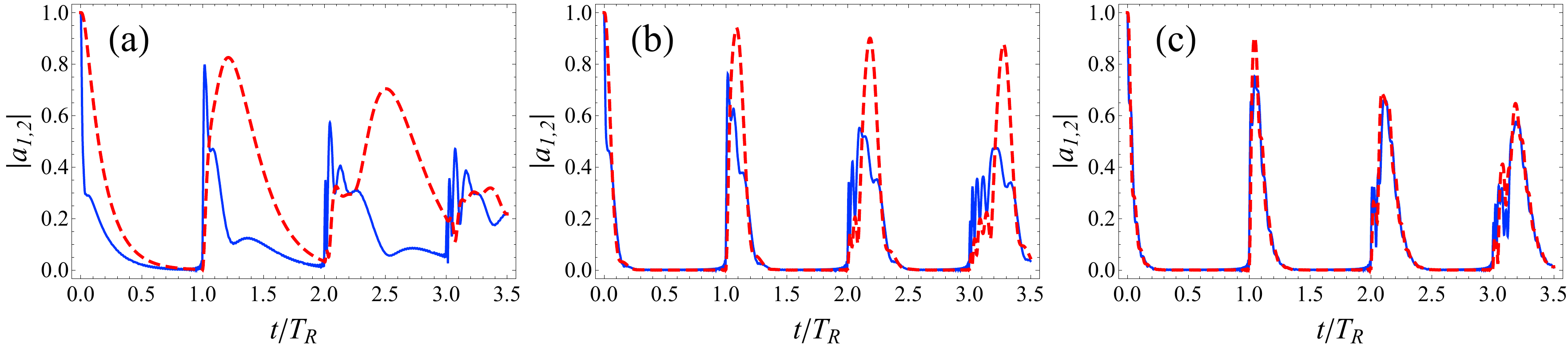}
\caption{The dependence of $\left| {{a_1}\left( t \right)} \right|$ (solid blue line) and $\left| {{a_2}\left( t \right)} \right|$ (dashed red line) on time at the coupling strengths $\Omega  = 0.5{\Omega _{SSE}}$ (a), $\Omega  = {\Omega _{SSE}}$ (b), $\Omega  = 2{\Omega _{SSE}}$ (c). Here ${T_R} = 2\pi /\delta \omega = 2\pi \cdot 10^{3} \omega_0^{-1}$ ($\delta\omega=0.001\omega_0$ for $N=200$) is the return time; the number of oscillators in the set is $200$. The initial state is ${a_1}\left( 0 \right) = 1$; ${a_2}\left( 0 \right) = 1$; ${b_k}\left( 0 \right) = 0$. $\Omega_{SSE}=\gamma/\sqrt{2}$ and $\gamma  = \pi {g^2}/\delta \omega \approx 0.014\omega_0$.}
\label{fig4}
\end{figure*}

The non-Hermitian description is only approximate, since it does not explicitly describe the behavior of the oscillators in the set. Within the Born-Markovian approximation used by derivation of the non-Hermitian equation, it is considered that the state of the oscillators in the set does not depend on time and always coincides with the initial state. At the initial moment, the amplitudes of oscillators in the set are zero and, therefore, there is only energy flow from the first and second oscillators to the set of oscillators. The energy flow in the opposite direction is absent. In this case, exponential decay of the amplitudes of the first and second oscillators is observed. This description ceases to work at $t > {T_R} = 2\pi /\delta \omega $, when a strong reverse flow is formed in the set due to interference between the amplitudes of the oscillators \cite{ref42N}. Such a flow leads to the appearance of revivals that are repeated at an interval ${T_R}$ [Figure~\ref{fig4}] \cite{ref42N,sergeev2022signature,ferreira2021collapse}. In this case, the dynamics of the system cannot be described using non-Hermitian equations~(\ref{eq:refname5}). A complete description of the system is given by the Hermitian equations~(\ref{eq:refname2})-(\ref{eq:refname4}). However, there is always a set of eigenstates of the Hermitian system that have the same symmetries as the Hamiltonian of the system. Therefore, spontaneous symmetry breaking cannot occur in the eigenstates of the Hermitian system. At the same time, the transitions with the change in symmetry are observed in the behavior of active/dissipative systems \cite{ref13,ref14}, which actually are only subsystems of the Hermitian systems. Therefore, the question arises: what causes such transitions in Hermitian systems? Are the transitions observed only at small time ($t < {T_R}$) when the non-Hermitian description is applicable? And does PT-symmetry in active/dissipative systems arise only as a result of inaccurate description?

In the following, we demonstrate that although the Hermitian system and its eigenstates are non-symmetrical for all values of $\Omega$, there is a transition in which symmetry appears in the system dynamics. This transition takes place both for times smaller than ${T_R}$ and for times greater than ${T_R}$.

\section{Change in eigenstates of the Hermitian system}
Non-Hermitian equations correctly describe the system evolution only at $t <  < 2\pi /\delta \omega$. At longer times, $t \ge 2\pi /\delta \omega$, the Hermitian system demonstrates complex dynamics including collapses and revivals of the energy oscillations~\cite{ref40,ref41,ref42N,ref47}, which arise at times greater than the return time ${T_R} = 2\pi /\delta \omega$ \cite{ref42N}. On this timescale, the system evolution is no longer described by the non-Hermitian equation~(\ref{eq:refname5}) and is determined by the eigenfrequencies ${f_k}$ and the eigenstates ${{\bf{e}}_k} = {({e_k})_j}$ ($j,k \in 1,...,N + 2$) of the matrix on the right side of Eqns.~(\ref{eq:refname2})-(\ref{eq:refname4}).

\begin{figure}[htbp]
\centering\includegraphics[width=\linewidth]{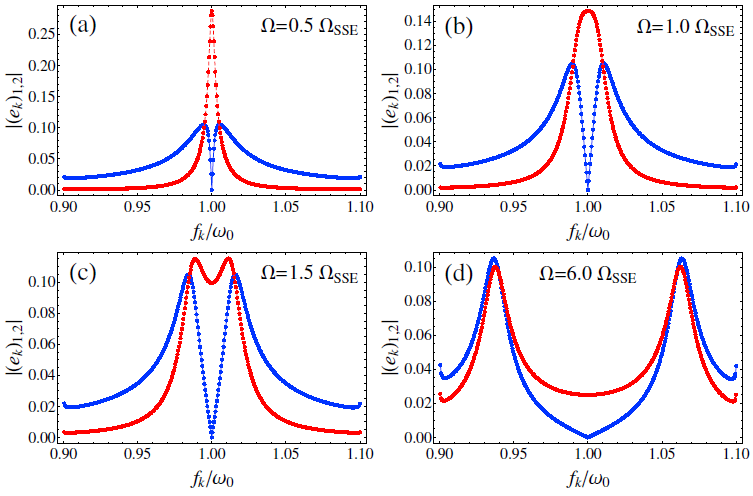}
\caption{The dependence of the amplitudes of the first ${({e_k})_1}$ (blue points) and second ${({e_k})_2}$ (red points) components in the eigenstate ${{\bf{e}}_k}$ on its eigenfrequency ${f_k}$ for the different coupling strengths: $\Omega  = 0.5\,{\Omega _{SSE}}$ (a); $\Omega  = \,{\Omega _{SSE}}$ (b); $\Omega  = 1.5\,{\Omega _{SSE}}$ (c); $\Omega  = 6.0\,{\Omega _{SSE}}$ (d). Here ${\Omega _{SSE}} = \gamma /\sqrt 2$, $\gamma  = \pi {g^2}/\delta \omega \approx 0.014\omega_0$, $\delta\omega = 0.0005\omega_0$, $g=0.0015\omega_0$ (for $N=400$) and we consider $\omega_0=1$.. The figures are plotted for $N=400$.}
\label{fig2}
\end{figure}

The considered system~(\ref{eq:refname2})-(\ref{eq:refname4}) is Hermitian; thus, its eigenfrequencies are real and all eigenstates are mutually orthogonal. The variation of the coupling strength $\Omega $ leads to a change in the amplitudes of the components of the different eigenstates. We trace out the dependence of the amplitudes of the first and second components, ${({e_k})_1}$ and ${({e_k})_2}$, of the eigenstate ${{\bf{e}}_k}$ on its eigenfrequency ${f_k}$ [Figure~\ref{fig2}]. These components correspond to the contributions of the first and second oscillators to the eigenstate ${{\bf{e}}_k}$. It is seen that the eigenstates do not possess symmetry with respect to the permutation of the first and second oscillators ($\left| {{{\left( {{{\bf{e}}_k }} \right)}_1}} \right| \ne \left| {{{\left( {{{\bf{e}}_k }} \right)}_2}} \right|$) at any values of $\Omega$. This is due to the fact that the Hermitian system~(\ref{eq:refname2})-(\ref{eq:refname4}) is not symmetrical and the eigenstates of Hermitian systems always possess the same symmetry as the system. However, an increase in the coupling strength is still accompanied by a qualitative change in the behaviour of the eigenstates components.

At small values of the coupling strength, the value of the second component of the eigenstate with eigenfrequency ${f_k} = {\omega _0}$ is larger than the value of the second component of any other eigenstates, see the peak of the red curve in Figures~\ref{fig2}(a), (b). In contrast, this eigenstate with ${f_k} = {\omega _0}$ has a zero value of the first component [blue curves in Figures~\ref{fig2}(a), (b)]. The increase in coupling strength leads to a splitting of the peak in the dependence of the second components of the eigenstates on their eigenfrequencies. From the numerical calculations of the eigenstates, we obtain that the splitting begins to appear at the critical coupling strength $\Omega  = {\Omega _{SSE}} = \gamma /\sqrt 2$ [Figure~\ref{fig2}(b), the red curve]. At $\Omega  > {\Omega _{SSE}}$, some eigenstates with eigenfrequencies ${f_k} \ne {\omega _0}$ have the highest value of the second components [the red curve in Figure~\ref{fig2}(c)]. This reflects the fact that it is not the individual first oscillator that interacts with the set of oscillators but rather the coherent superposition of the first and second oscillators. At an increase in coupling strength above the critical value, the dependencies of the first and second components of the eigenstates on the eigenfrequencies ${f_k}$ begin to become similar to each other [cf. the red and blue curves in Figures~\ref{fig2}(c), (d)]. Thus, the qualitative change in the eigenstates occurs at $\Omega  = {\Omega _{SSE}}$. When $N >  > 1$, such a change in the eigenstates of the Hermitian system does not depend on $N$ (see Appendix E). Moreover, this change occurs even when the number of oscillators is about $10$. That is, the change in the eigenstates is a general phenomenon in such Hermitian systems.

The critical coupling strength ${\Omega _{SSE}} = \gamma /\sqrt 2$ does not coincide with the coupling strength corresponding to the EP (${\Omega _{EP}} = \gamma /2$), at which the splitting in the non-Hermitian system~(\ref{eq:refname5}) appears. However, the coupling strength corresponding to the EP is calculated without taking into account the noise terms. The presence of noise leads to the fact that splitting in the spectrum of the second oscillators occurs at the coupling strength $\Omega= \gamma / \sqrt 2$ (see Appendix C and Ref.~\cite{ref15}) that coincides with $\Omega_{SSE}$. Therefore, the splitting in the second components  ${({e_k})_2}$ can be associated with the splitting in the spectrum of the stochastic non-Hermitian system.

\section{Spontaneous symmetry emergence}
The change in eigenstates manifests itself in the temporal dynamics of the Hermitian system~(\ref{eq:refname2})-(\ref{eq:refname4}). To illustrate this fact, we calculate the temporal dynamics of the oscillators [Figure~\ref{fig4}] using the obtained eigenstates and eigenfrequencies of Eqns.~(\ref{eq:refname2})-(\ref{eq:refname4}). When $\Omega  < {\Omega _{SSE}}$, the absolute value of the amplitude of the first oscillator is smaller than the one of the second oscillator [Figure~\ref{fig4}(a)]. It is due to the fact that the first oscillator interacts with the set of oscillators, while the second oscillator does not. With increasing coupling strength, the dynamics of the oscillators' amplitudes become similar to each other [Figures~\ref{fig4}(b), (c)]. At the same time, the phase difference between the oscillators changes over time for all values of $\Omega $ (see Figure~\ref{fig1A} in Appendix D). That is, there is no phase synchronization \cite{ref48} in the system.

To trace out the changes in the oscillators' amplitudes, we calculate the ratio of the absolute values of the amplitudes of the first and second oscillators averaged over time ${\left\langle {...} \right\rangle _t}$. To begin with, we choose ${a_1}\left( {t = 0} \right) = {\left( {{{\bf{e}}_ + }} \right)_1}$, ${a_2}\left( {t = 0} \right) = {\left( {{{\bf{e}}_ + }} \right)_2}$ (see Eq.~(\ref{eq:refname1N})), and ${b_k}\left( {t = 0} \right) = 0$  as the initial state. When the averaging is carried out over a time interval before the first revival ($0 \le t < {T_R}$), the ratio ${\left\langle {\left| {{a_1}} \right|} \right\rangle _t}/{\left\langle {\left| {{a_2}} \right|} \right\rangle _t}$ demonstrates the threshold-like behavior at $\Omega  = {\Omega _{EP}}$  [Figure~\ref{fig3}]. This behavior is similar to the one observed in the non-Hermitian equation~(\ref{eq:refname5}) at spontaneous symmetry breaking. This similarity is due to the fact that the dynamics of Hermitian system at times $t<T_R$ is like the one of non-Hermitian system~(\ref{eq:refname5}) [Figure~\ref{fig4}]. However, at $t > {T_R}$ the behavior of ${a_1}\left( t \right)$ and ${a_2}\left( t \right)$ in the Hermitian system differs qualitatively from the one in the non-Hermitian system [Figure~\ref{fig4}]. When  ${a_1}\left( t \right)$ and ${a_2}\left( t \right)$ are averaged over a time interval containing a large number of collapses and revivals ($0 \le t < {t_{\max }}$, where ${t_{\max }} >  > {T_R}$), the ratio ${\left\langle {\left| {{a_1}} \right|} \right\rangle _t}/{\left\langle {\left| {{a_2}} \right|} \right\rangle _t}$ demonstrates the threshold-like behavior at $\Omega  = {\Omega _{SSE}}$ [Figure~\ref{fig3}]. Thus, it is seen that in the Hermitian system, the symmetry appears at $\Omega  = {\Omega _{SSE}}$. For $t> >T_R$, the curve ${\left\langle {\left| {{a_1}} \right|} \right\rangle _t}/{\left\langle {\left| {{a_2}} \right|} \right\rangle _t}$ is rough, which is due to the fact that after a large number of revivals the temporal dynamics of the amplitudes is quasi-random [Figure~\ref{fig4}]. However, the threshold is clearly visible in the dependence of ${\left\langle {\left| {{a_1}} \right|} \right\rangle _t}/{\left\langle {\left| {{a_2}} \right|} \right\rangle _t}$ on $\Omega$.

\begin{figure}[htbp]
\centering\includegraphics[width=0.7\linewidth]{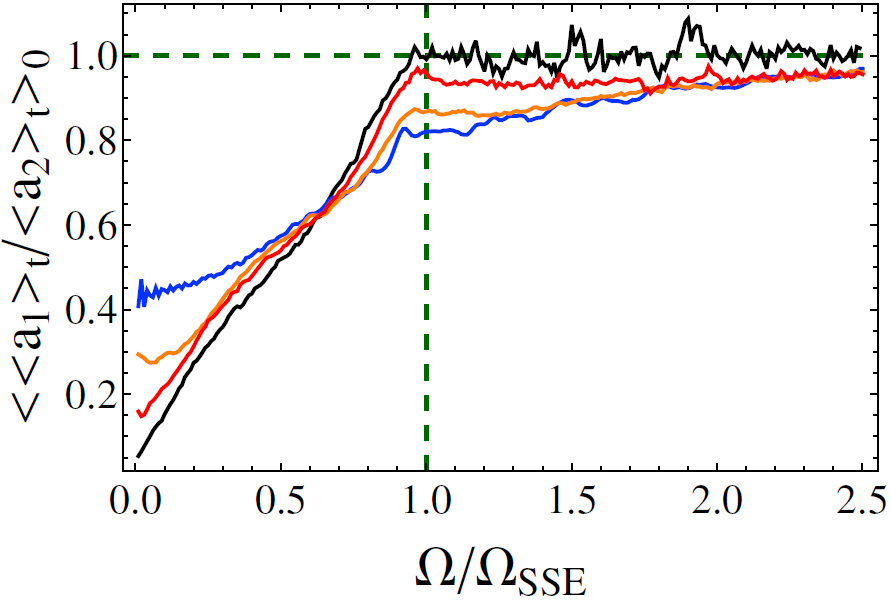}
\caption{The dependence of ${\left\langle {{{\left\langle {|{a_1}|} \right\rangle }_t}/{{\left\langle {|{a_2}|} \right\rangle }_t}} \right\rangle _0}$ on the coupling strength, $\Omega$, when the number of oscillators in the set is $N=50$ (the blue line), $N=100$ (the orange line), $N=200$ (the red line), $N=400$ (the black line). The dashed vertical line shows the critical coupling strength ${\Omega _{SSE}} = \gamma /\sqrt 2 $. The dashed horizontal line shows $1$. The interval of averaging is $\left[ {0,\,25\,{T_R}} \right]$. The averaging is carried out over 200 initial states.}
\label{fig5}
\end{figure}

The Eqns.~(\ref{eq:refname2})-(\ref{eq:refname4}) for the Hermitian system do not possess symmetry with respect to the permutation of the first and second oscillators. That is, the symmetry emerges in the system, which has no such symmetry. Similar behavior takes place for other initial states. To illustrate this fact, we calculate the ratio of the absolute values of the amplitudes of the first and second oscillators averaged over time and then over the initial states, ${\left\langle {{{\left\langle {|{a_1}|} \right\rangle }_t}/{{\left\langle {|{a_2}|} \right\rangle }_t}} \right\rangle _0}$ [Figure~\ref{fig5}]. Our calculations show that when the averaging is carried out over a time interval before the first revival ($0 \le t < {T_R}$), the ratio ${\left\langle {{{\left\langle {\left| {{a_1}} \right|} \right\rangle }_t}/{{\left\langle {\left| {{a_2}} \right|} \right\rangle }_t}} \right\rangle _0}$ increases smoothly with the coupling strength $\Omega $ (see Figure~\ref{fig2A} in Appendix D). Instead, when the averaging is carried out over a time interval containing a large number of the collapses and revivals ($0 \le t < {t_{\max }}$, where ${t_{\max }} >  > {T_R}$), the ratio ${\left\langle {{{\left\langle {\left| {{a_1}} \right|} \right\rangle }_t}/{{\left\langle {\left| {{a_2}} \right|} \right\rangle }_t}} \right\rangle _0}$ demonstrates the threshold-like behavior at $\Omega  = {\Omega _{SSE}}$ [Figure~\ref{fig5}]. 

\section{Dependence of the system behavior on the number of oscillators in the set}
To prove the existence of the transition in the considered system, we study the dependence of ${\left\langle {{{\left\langle {\left| {{a_1}} \right|} \right\rangle }_t}/{{\left\langle {\left| {{a_2}} \right|} \right\rangle }_t}} \right\rangle _0}$ on the number of oscillators in the set. The frequencies of the oscillators in the set uniformly fill the interval from $-N \delta \omega/2$  to $N \delta \omega/2$. To keep the frequency interval unchanged, we scale $\delta \omega$ as $N^{-1}$ ($N \delta \omega = const$, that is, we increase the density of states in the set of the oscillators). The increase in the density of states leads to an enhancement of the energy exchange between the first oscillator and the set of oscillators. Based on the expression for the relaxation rate, $\gamma  = \pi {g^2}/\delta \omega$, (\ref{eq:10A}), we conclude that such an enhancement is proportional to $1/\delta \omega  \sim N$. To keep the magnitude of the interaction of the first oscillator with the set of oscillators constant, we scale $g$ as $1/\sqrt N$ (see Appendix E). Our calculations show that such a scaling preserves the behavior of the eigenstates of the Hermitian system (see Figure~\ref{fig3A} in Appendix E). The change of $\delta \omega$ causes the change in the return time (${T_R} = 2\pi /\delta \omega$). To keep the number of revivals and collapses occurring during the observation time, we scale the observation time as $N$.

Our calculations show that when the averaging of ${\left\langle {{{\left\langle {\left| {{a_1}} \right|} \right\rangle }_t}/{{\left\langle {\left| {{a_2}} \right|} \right\rangle }_t}} \right\rangle _0}$ is carried out over a time interval containing a large number of collapses and revivals, the increase in $N$ makes the transition at $\Omega  = {\Omega _{SSE}}$ sharper [Figure~\ref{fig5}]. In the limit of an infinite set ($\delta \omega /{\omega _0} \to 0$ and $N \to \infty$), at $\Omega  > {\Omega _{SSE}}$ the ratio ${\left\langle {{{\left\langle {\left| {{a_1}} \right|} \right\rangle }_t}/{{\left\langle {\left| {{a_2}} \right|} \right\rangle }_t}} \right\rangle _0}$ tends to $1$ [Figure~\ref{fig5}]. Thus, the oscillators' amplitudes become equal to each other, and in such a sense, the system dynamics become symmetric with respect to the permutation of the oscillators. Note that the system~(\ref{eq:refname1}) is not symmetric with respect to the permutation of the oscillators. That is, the symmetry spontaneously emerges in the system dynamics, leading to the threshold change in the dependence of ${\left\langle {{{\left\langle {|{a_1}|} \right\rangle }_t}/{{\left\langle {|{a_2}|} \right\rangle }_t}} \right\rangle _0}$ on the coupling strength [Figure~\ref{fig5}]. We name this phenomenon as a spontaneous symmetry emergence. The spontaneous symmetry emergence can be considered as a new class of phase transitions connected with a change in symmetry of the system states.

Note that when the averaging of ${\left\langle {{{\left\langle {\left| {{a_1}} \right|} \right\rangle }_t}/{{\left\langle {\left| {{a_2}} \right|} \right\rangle }_t}} \right\rangle _0}$ is carried out over the time interval smaller than the time of first revival, the increase of the number of the oscillators in the set ($\delta \omega /{\omega _0} \to 0$ and $N \to \infty$) does not lead to an appearance of the threshold (see Figure~\ref{fig2A} in Appendix D). Thus, the spontaneous symmetry emergence is observed only when the averaging is carried out over a long time interval ($t >  > {T_R}$), during which a large number of collapses and revivals take place [Figure~\ref{fig5}]. It indicates that the spontaneous symmetry emergence is the property of the entire Hermitian system~(\ref{eq:refname1}) and cannot be associated with the properties of the non-Hermitian system~(\ref{eq:refname5}) obtained by the elimination of the degrees of freedom of the set of oscillators.

In our work, we study symmetry based on the ratio of time-averaged oscillator amplitudes (${\left\langle {{{\left\langle {\left| {{a_1}} \right|} \right\rangle }_t}/{{\left\langle {\left| {{a_2}} \right|} \right\rangle }_t}} \right\rangle _0}$). The transition with the appearance of symmetry that we predicted manifests itself averaging $\left| {{a_1}} \right|$ and $\left| {{a_2}} \right|$ over a long time interval. Although at an arbitrary moment of time $\left| {{a_1}} \right| \ne \left| {{a_2}} \right|$. The use of time averaging is applicable in determining symmetry in a finite-size system. In fact, in our system, the set of oscillators plays the role of a finite-sized reservoir. The interaction of the first and second oscillators with the reservoir leads to direct and reverse energy flows between them, which makes the dynamics of the system complex (chaotic-like) at large times. This behavior is associated with the finite size of the system under consideration, and the definition of symmetry that we use is similar to that used in statistical physics for finite-sized systems. According to the Ginzburg-Landau theory \cite{ref1}, the order parameter is zero in the symmetric state and becomes non-zero after the phase transition point. However, due to the interaction with the environment, the order parameter fluctuates in time \cite{ref1}. The zero value of the order parameter is obtained by averaging over an observation time. In a finite-size sample, the magnitude of fluctuations can be comparable to the average value of the order parameter even in a non-symmetric phase (at some region near the phase transition point). Therefore, to establish phase symmetry and the transition point, it is necessary to perform time averaging. Thus, our definition of symmetry is similar to the definition in statistical physics for finite-sized systems. We ignore the difference in the values of $\left| {{a_1}} \right|$ and $\left| {{a_2}} \right|$ at each individual moment of time and consider only the averaged values. This approach is used for finite-sized systems when the fluctuations play an important role. 

It is important that the curve ${\left\langle {{a_1}} \right\rangle _t}/{\left\langle {{a_2}} \right\rangle _t}$ depends on whether the averaging is carried out over a time less than $T_R$ or over a time much greater than $T_R$ [Figure~\ref{fig3}]. In particular, $\Omega_{EP} \ne \Omega_{SSE}$. There is a fundamental difference between case when the averaging time $t < {T_R}$ and case when the averaging time $t > > {T_R}$. In Figure~\ref{fig3} one of the eigenstates of the non-Hermitian equation~(\ref{eq:refname5}) is selected as the initial state. At times up to the first revival, the oscillator amplitudes decay exponentially at a rate determined by the imaginary part of the eigenvalue of the corresponding eigenstate. As a result, the system remains near the initial eigenstate (only the amplitude of the state decreases). At larger times, revivals occur in the Hermitian system, which lead to transitions to other states. The difference in the transition points from the non-symmetric state to the symmetric one at small and large averaging times is due to the fact that at times before the first revival the system does not have time to transition from the initial eigenstate, which is symmetric at $\Omega  > {\Omega _{EP}}$ (see Eq.~(\ref{eq:refname1N})). When averaging over larger times, the system has time to deviate from the initial state, which leads to a change in the transition point from $\Omega  = {\Omega _{EP}}$ to $\Omega  = {\Omega _{SSE}} > {\Omega _{EP}}$. It is important that with an additional averaging over the initial conditions, the transition point obtained by averaging over a long period of time does not change ($\Omega  = {\Omega _{SSE}}$) [Figure~\ref{fig5}]. At the same time, the transition point obtained by averaging over a short time disappears with additional averaging over the initial conditions (that is, the curve ${\left\langle {{{\left\langle {{a_1}} \right\rangle }_t}/{{\left\langle {{a_2}} \right\rangle }_t}} \right\rangle _0}$ is smooth) [Figure~\ref{fig2A}b].

Thus, we conclude that the transition at the exceptional point predicted for the non-Hermitian system is approximate and can be observed in the Hermitian systems only at times before the first revival and when the initial states coincided with the eigenstates. The transition to the symmetric state in the Hermitian system occurs at large coupling strengths.

\section{Discussion of a mechanism of spontaneous symmetry emergence}
The symmetry with respect to the permutation of the oscillators begins to emerge at the coupling strength $\Omega  = {\Omega _{SSE}}$. It is the coupling strength at which the change in eigenstates occurs. Below the critical coupling, the eigenstates ${{\bf{e}}_k}$ are not symmetrical with respect to the permutation of the absolute value of the first and second components [Figure~\ref{fig2}(a)], and the first and second components of the vast majority of eigenstates differ significantly from each other [Figure~\ref{fig2}(a)]. At the critical coupling strength, the peak in the dependence of second components of eigenstates ${{\bf{e}}_k}$ on their eigenfrequencies ${f_k}$ splits into two peaks [Figure~\ref{fig2}(b)]. At further increases in the coupling strength, the eigenstates begin to become symmetrical with respect to the permutation of the first and second components [Figure~\ref{fig2}(b), (c), (d)]. 

The transformation of the eigenstates causes the change in the oscillators' amplitudes. The main contribution to the amplitudes of the first and second oscillators is given by the eigenstates, which have the greatest values of the corresponding components. Therefore, below the critical coupling strength ($\Omega  < {\Omega _{SSE}}$), the main contribution to the amplitude of the second oscillator ${\left\langle {|{a_2}|} \right\rangle _t}$ is given by the eigenstates, the eigenfrequencies of which lie within the width of peak at ${f_k} = {\omega _0}$ [the red line in Figure~\ref{fig2}(a)]. At the same time, the main contribution to the amplitude of the first oscillator ${\left\langle {|{a_1}|} \right\rangle _t}$ is given by the eigenstates, the eigenfrequencies of which lie within the widths of two peaks at ${f_k} \ne {\omega _0}$ [the blue line in Figure~\ref{fig2}(a)]. The numbers of eigenstates, giving the main contribution to ${\left\langle {|{a_1}|} \right\rangle _t}$, depend on $\Omega$, while the numbers of eigenstates, giving the main contribution to ${\left\langle {|{a_2}|} \right\rangle _t}$, do not. Above the critical coupling strength ($\Omega  > {\Omega _{SSE}}$), the main contributions to the amplitudes of both oscillators come from the eigenstates, the eigenfrequencies of which lie within the line widths of two peaks at ${f_k} \ne {\omega _0}$ [Figures~\ref{fig2}(c), (d)]. Moreover, at $\Omega  > {\Omega _{SSE}}$, the numbers of eigenstates, giving the main contributions to both ${\left\langle {|{a_1}|} \right\rangle _t}$ and ${\left\langle {|{a_2}|} \right\rangle _t}$, depend on $\Omega$. Thus, the dependence of ${\left\langle {|{a_2}|} \right\rangle _t}$ on the coupling strength undergoes qualitative change, which manifests itself in the threshold behavior of ${\left\langle {|{a_1}|} \right\rangle _t}/{\left\langle {|{a_2}|} \right\rangle _t}$ [Figure~\ref{fig5}]. With the increase in coupling strength, the first and second components of eigenstates become similar to each other [Figure~\ref{fig2}(d)]. Our analytical estimations show that (see Appendix F) the ratio of ${\left\langle {|{a_1}|} \right\rangle _t}/{\left\langle {|{a_2}|} \right\rangle _t}$ is determined by the ratio of ${P_1}\sqrt {{\Gamma _1}} /{P_2}\sqrt {{\Gamma _2}}$, where ${P_{1,2}} = \max \left( {{e_k}} \right)_{1,2}^2$ are the heights of peaks, $\Gamma _{1,2}$ are the widths of the peaks (if there are two peaks, then $\Gamma _{1,2}$ are the sum of the widths of both the peaks). This ratio does not depend on the peak frequencies. Above the critical point, the ratio of ${P_1}\sqrt {{\Gamma _1}} /{P_2}\sqrt {{\Gamma _2}}$ tends to 1 (see Appendix F for details) and, therefore, the ratio of ${\left\langle {|{a_1}|} \right\rangle _t}/{\left\langle {|{a_2}|} \right\rangle _t}$ tends to $1$ too. Thus, the emergence of the symmetry in the system dynamics can be unambiguously associated with the changes in the eigenstates.

\section{Conclusion}
We predict the existence of a new class of phase transitions that relate to the emergence of symmetry in the state of the Hermitian system, the Hamiltonian of which does not possess such a symmetry. In other words, the symmetry emerges spontaneously in the system's behavior. This phenomenon is the counterpart of spontaneous symmetry breaking, which is a conventional concomitant of second-order phase transitions. From the point of view of an external observer, spontaneous symmetry emergence has the same manifestations as spontaneous symmetry breaking. Indeed, in both cases, the symmetry of the system state changes at the transition point. However, in contrast to the case of spontaneous symmetry breaking, in which the system Hamiltonian has the same symmetry as the state in symmetrical phase, in the case of spontaneous symmetry emergence, the system Hamiltonian does not have the emerged symmetry.

The spontaneous symmetry emergence is the source of a new class of phase transitions with a change in symmetry; for example, in an optical system consisting of two coupled single-mode cavities, one of which interacts with a waveguide of finite length. Thus, our results open the way to the study of phase transitions with the change of the symmetry in Hermitian systems without symmetry.

\section*{Acknowledgement}
T.T.S. and E.S.A. thanks foundation for the advancement of theoretical physics and mathematics “Basis”.

\bibliographystyle{plain}

\section*{Appendix A. Derivation of the systems’ equations}
We consider the system of two coupled oscillators, one of which interacts with a finite set of oscillators. The Hamiltonian of the system is
\begin{equation}
\begin{gathered}
  \hat H = {\omega _0}\hat a_1^\dag {{\hat a}_1} + {\omega _0}\hat a_2^\dag {{\hat a}_2} + \Omega \left( {\hat a_1^\dag {{\hat a}_2} + \hat a_2^\dag {{\hat a}_1}} \right) +  \hfill \\
  \sum\limits_{k = 1}^N {{\omega _k}\hat b_k^\dag {{\hat b}_k}}  + \sum\limits_{k = 1}^N {g\left( {\hat b_k^\dag {{\hat a}_1} + \hat a_1^\dag {{\hat b}_k}} \right)}  \hfill \\ 
\end{gathered}
\label{eq:1A}
\end{equation}
where ${\hat a_{1,2}}$ and $\hat a_{1,2}^\dag$ are the annihilation and creation operators of the first and second oscillators, $\left[ {{{\hat a}_j},\hat a_k^\dag } \right] = {\delta _{jk}}$ \cite{ref3}. ${\hat b_k}$ and $\hat b_k^\dag$ are the annihilation and creation operators of the oscillators in the set with the frequencies ${\omega _k} = {\omega _0} + \delta \omega \left( {k - \frac{N}{2}} \right)$, $\left[ {{{\hat b}_j},\hat b_k^\dag } \right] = {\delta _{jk}}$. $\Omega$ is a coupling strength between the first and second oscillators. $g$ is a coupling strength between the first oscillator and each of the oscillators in the set. For convenience, we consider that $\hbar  = 1$.
Using the Heisenberg equation for operators $\frac{{d\hat A}}{{dt}} = i\left[ {\hat H,\hat A} \right]$ \cite{ref3}, we obtain the following equations:
\begin{equation}
\begin{gathered}
  \frac{{d{{\hat a}_1}}}{{dt}} = i\left[ {\hat H,{{\hat a}_1}} \right] =  - i{\omega _0}{{\hat a}_1} - i\Omega {{\hat a}_2} - i\sum\limits_{k = 1}^N {g{{\hat b}_k}}  \hfill \\
  \frac{{d{{\hat a}_2}}}{{dt}} = i\left[ {\hat H,{{\hat a}_2}} \right] =  - i{\omega _0}{{\hat a}_2} - i\Omega {{\hat a}_1} \hfill \\
  \frac{{d{{\hat b}_k}}}{{dt}} = i\left[ {\hat H,{{\hat b}_k}} \right] =  - i{\omega _k}{{\hat b}_k} - ig{{\hat a}_1} \hfill \\ 
\end{gathered}
\label{eq:2A}
\end{equation}
One can notice that the system~(\ref{eq:2A}) is linear and then we can move from equations for operators to closed system of equations for their averages:
\begin{equation}
\frac{{d{a_1}}}{{dt}} =  - i{\omega _0}{a_1} - i\Omega {a_2} - i\sum\limits_{k = 1}^N {g{b_k}}
\label{eq:3A}
\end{equation}

\begin{equation}
\frac{{d{a_2}}}{{dt}} =  - i{\omega _0}{a_2} - i\Omega {a_1}
\label{eq:4A}
\end{equation}

\begin{equation}
\frac{{d{b_k}}}{{dt}} =  - i{\omega _k}{b_k} - ig{a_1}
\label{eq:5A}
\end{equation}

Here we use the notations $\left\langle {{{\hat a}_1}} \right\rangle  = {a_1}$, $\left\langle {{{\hat a}_2}} \right\rangle  = {a_2}$ and $\left\langle {{{\hat b}_k}} \right\rangle  = {b_k}$. Note that since the operators' equations do not contain the products of the operators (e.g. ${\hat a_1}\,{\hat a_2}$) then the transition to the operators’ averages is exact.

\section*{Appendix B. Derivation of the equations for non-Hermitian system}
The transition to a non-Hermitian system can be carried out by increasing the number of modes in the reservoir to infinity with simultaneous decreasing the distance between the neighboring modes to zero ($N \to \infty$ and $\delta \omega /{\omega _0} \to 0$). In this limit, the characteristic time of the revival of the oscillations (${T_R} = \frac{{2\pi }}{{\delta \omega }}$) \cite{ref42N} is much longer than the observation time. Therefore, the system dynamics are described only by the exponential decay stage, as it takes place in the non-Hermitian system.

To derive the non-Hermitian equations describing the system dynamics during the exponential stage, we first formally integrate the equation~(\ref{eq:5A})
\begin{equation}
{b_k} = {b_k}(0){e^{ - i{\omega _k}t}} - ig\int\limits_0^t {d\tau {a_1}(\tau ){e^{ - i{\omega _k}(t - \tau )}}}
\label{eq:6A}
\end{equation}
After substituting the Eq.~(\ref{eq:6A}) into the Eq.~(\ref{eq:3A}), we obtain
\begin{equation}
{\dot a_1} =  - i{\omega _0}{a_1} - i\Omega {a_2} - \sum\limits_k {{g^2}\int\limits_0^t {d\tau {a_1}(\tau ){e^{ - i{\omega _k}(t - \tau )}}} }
\label{eq:7A}
\end{equation}
Here we assume that ${b_k}(0) = 0$. Note that such an initial state we use in all our calculations.

Using the Born-Markovian approximation \cite{ref38,ref39} and Sokhotskii-Plemelj formula \cite{plemelj1964problems}, one can calculate the integral in the equation~(\ref{eq:7A}) and obtain the following equations for non-Hermitian system \cite{ref38,sergeev2021environment}:
\begin{equation}
\begin{gathered}
  {{\dot a}_1} = ( - i{\omega _0} - \gamma ){a_1} - i\Omega {a_2} \hfill \\
  {{\dot a}_2} =  - i{\omega _0}{a_2} - i\Omega {a_1} \hfill \\ 
\end{gathered}
\label{eq:8A}
\end{equation}
Here the decay rate $\gamma$ in~(\ref{eq:8A}) is determined by the following formula \cite{ref38}
\begin{equation}
\gamma  = \sum\limits_k {\pi {g^2}\delta \left( {{\omega _k} - {\omega _0}} \right)} 
\label{eq:9A}
\end{equation}
where $\delta (\omega )$ is Dirac’s delta-function.

The formula~\ref{eq:9A} can be simplified by substituting ${\omega _k} = {\omega _0} + \delta \omega \left( {k - N/2} \right)$. Then we obtain 
\begin{equation}
\gamma  = \frac{{\pi {g^2}}}{{\delta \omega }}
\label{eq:10A}
\end{equation}

The equation~(\ref{eq:refname5}) coincides with the equation for the average values of the annihilation operators of the first and second oscillators, derived from the density matrix of two coupled oscillators in the Lindblad form. The Lindblad equations are derived within the Born-Markovian approximation \cite{ref38}. In this approach, the degrees of freedom of the corresponding to the oscillators in the set (the reservoirs' degrees of freedom) are eliminated from the consideration. As a result, the master equation for the density matrix \cite{ref38} of two coupled oscillators is obtained:

\begin{equation}
\frac{d \hat{\rho}}{d t}= i \left[ \hat{H}_{12}, \hat{\rho} \right] + \frac{\gamma}{2} \left( 2 \hat{a}_2 \hat{\rho} \hat{a^\dag_2} - \hat{a}_2 \hat{a}^\dag_2 \hat{\rho} - \hat{\rho} \hat{a}_2 \hat{a}^\dag_2\right)
\label{eq:11A}
\end{equation}
where $\hat{H}_{12} = \omega_0 \hat{a}^\dag_1 \hat{a}_1 + \omega_0 \hat{a}^\dag_2 \hat{a}_2 + \Omega \left( \hat{a}^\dag_1 \hat{a}_2 + \hat{a}_1 \hat{a}^\dag_2  \right)$ is the Hamiltonian of two coupled oscillators and the second term in the right part of the Eq.~(\ref{eq:1A}).

Using the expressions $\left\langle {\hat{a}_{1,2}}  \right\rangle = Tr \left( \hat{\rho} \hat{a}_{1,2}\right)$, we obtain the equation for the average values of the annihilation operators of the first and second oscillators. These equations coincide with the Eqns.~(\ref{eq:8A}).

\section*{Appendix C. Non-Hermitian equations with noise}
In Appendix B, we used the assumption that $b_k (0)=0$, so we excluded the thermal effects of the environment on the system from consideration. However, in order to consider the radiation of such a system, it is necessary to take into account the noise in the system. Now, we assume that $b_k (0) \neq 0$. Then equations~(\ref{eq:8A}) will take the form of equations with noise~(\ref{eq:refname5}):
\begin{equation}
\frac{d}{{dt}}\left( {\begin{array}{*{20}{c}}
{{a_1}}\\
{{a_2}}
\end{array}} \right) = \left( {\begin{array}{*{20}{c}}
{ - i{\omega _0} - \gamma }&{ - i\Omega }\\
{ - i\Omega }&{ - i{\omega _0}}
\end{array}} \right)\left( {\begin{array}{*{20}{c}}
{{a_1}}\\
{{a_2}}
\end{array}} \right) + \left( {\begin{array}{*{20}{c}}
{{\xi}}\\
{{0}}
\end{array}} \right)
\label{eq:1C}
\end{equation}
where $\xi (t) = -i \sum\limits_k {g b_k (0) e^{-i (\omega_k - \omega_0) t}}$, according to fluctuation-dissipation theorem \cite{ref37,ref38,ref39}, is the noise term with the following correlations:
\begin{equation}
\langle \xi (t) \rangle = 0
\label{eq:2C}
\end{equation}
\begin{equation}
\langle \xi^{*} (t+\tau) \xi (t) \rangle = \gamma T \delta (\tau)
\label{eq:3C}
\end{equation}
where $\gamma$ is determined by Eq.~(\ref{eq:10A}) and $T$ is the temperature of the environment. In this model it is possible to calculate the spectrum of the two main oscillators (see Eq. (6) in Ref.~\cite{ref15}). Then we can calculate the value of coupling strength, $\Omega_{2}^{split}$, when the splitting of the second oscillator's spectrum appears. To do this, we can use the formula from our work~\cite{ref15} that was obtained for more general models in which each oscillator interacts with its reservoir (see pp. 9 - 10 in Ref.~\cite{ref15}):
\begin{equation}
\begin{split}
(\Omega_{2}^{split})^{2}= \frac{\gamma_{2}^{2}\gamma_{1}T_{1}+\gamma_{1}^{3}T_{1}-2\gamma_{1}^{2}\gamma_{2}T_{2}-2\gamma_{2}^{2}\gamma_{1}T_{2}}{2(\gamma_{2}T_{2}+2\gamma_{1}T_{1})} + \\
\frac{\sqrt{4\gamma_{2}\gamma_{1}^{4}T_{2}(\gamma_{2}T_{2}+2\gamma_{1}T_{1})+(\gamma_{1}(\gamma_{1}^{2}+\gamma_{2}^{2})T_{1}-2\gamma_{1}\gamma_{2}(\gamma_{1}+\gamma_{2})T_{2})^{2}}}{2(\gamma_{2}T_{2}+2\gamma_{1}T_{1})}
\end{split}
\label{eq:4C}
\end{equation}

Here $\gamma_1$, $\gamma_2$ are the relaxation rates of the first and second oscillators. $T_1$ and $T_2$ are temperatures of the reservoirs interacting with the first and second oscillators, respectively.

Using the fact that the our system has only one reservoir, i.e. $\gamma_{1}=\gamma$, $\gamma_{2}=0$, $T_{1}=T$ and $T_{2}=0$, we finally obtain:
\begin{equation}
\Omega_{2}^{split} = \gamma/\sqrt{2}
\label{eq:5C}
\end{equation}
This value of coupling strength coincides with the critical coupling strength, above which the splitting of the eigenstates of the Hermitian system occurs [see red lines in Figure~\ref{fig2}], i.e. $\Omega_{2}^{split} = \Omega_{SSE} = \gamma/\sqrt{2}$. Note that the spectrum of the first oscillator is always splitted, which is also predicted by the considered model [see blue lines in Figure~\ref{fig2}].

\section*{Appendix D. Dynamics of the Hermitian system at different coupling strength between the oscillators}
\begin{figure}[htbp]
\centering\includegraphics[width=\linewidth]{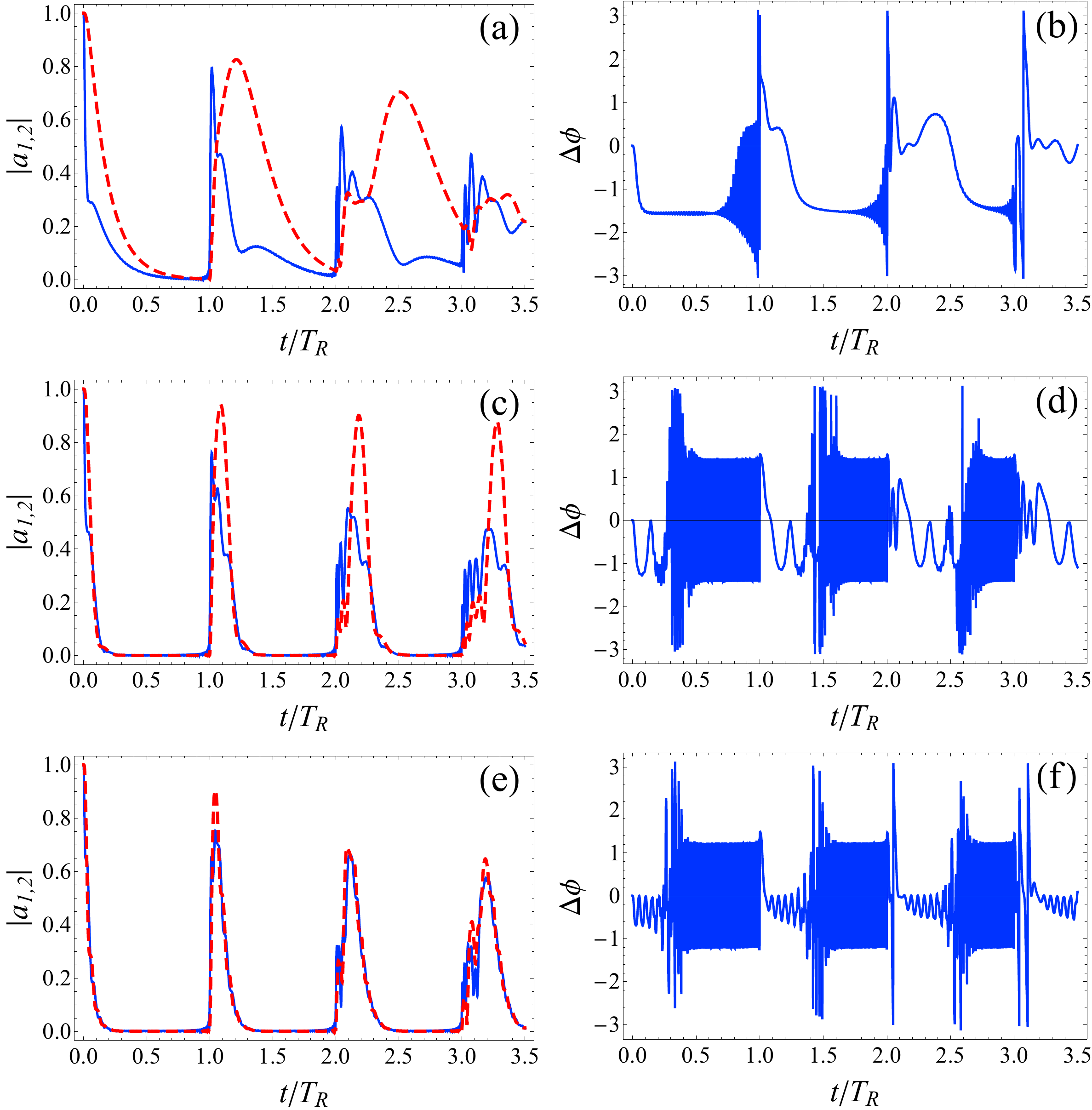}
\caption{(a), (c), (e) The dependence of $\left| {{a_1}\left( t \right)} \right|$ (solid blue line) and $\left| {{a_2}\left( t \right)} \right|$ (dashed red line) on time. (b), (d), (f) The dependence of the phase difference between the amplitudes of the first and second oscillators ($\Delta \phi  = \arg \left( {{a_1}\left( t \right)} \right) - \arg \left( {{a_2}\left( t \right)} \right)$) on time. The coupling strength between the oscillators is $\Omega  = 0.5\,{\Omega _{SSE}}$ (a), (b); $\Omega  = {\Omega _{SSE}}$ (c), (d); $\Omega  = 2\,{\Omega _{SSE}}$ (e), (f). Here ${T_R} = 2\pi /\delta \omega $ is a return time; the number of oscillators in the set is $200$. The initial state is  ${a_1}\left( 0 \right) = 1$; ${a_2}\left( 0 \right) = 1$; ${b_k}\left( 0 \right) = 0$.}
\label{fig1A}
\end{figure}

We use the Eqns.~(\ref{eq:3A})--(\ref{eq:5A}) to describe the dynamics of the Hermitian system~(\ref{eq:1A}). We calculate the temporal evolution of the oscillators’ amplitudes, $\left| {{a_{1,2}}\left( t \right)} \right|$, and the phase difference between the oscillators, $\Delta \phi \left( t \right) = \arg \left( {{a_1}\left( t \right)} \right) - \arg \left( {{a_2}\left( t \right)} \right)$ [Figure~\ref{fig1A}]. When the coupling strength between the oscillators $\Omega  <  < \,{\Omega _{SSE}}$, the temporal dynamics of $\left| {{a_1}\left( t \right)} \right|$  and $\left| {{a_2}\left( t \right)} \right|$ are very different from each other [Figure~\ref{fig1A}(a)]. Increasing the coupling strength causes the temporal dynamics of $\left| {{a_1}\left( t \right)} \right|$ and $\left| {{a_2}\left( t \right)} \right|$ to become more similar [Figure~\ref{fig1A}(c) and (e)]. In turn, when $\Omega  <  < \,{\Omega _{SSE}}$, the phase difference between the oscillators, $\Delta \phi \left( t \right)$, changes noticeably during the collapses and revivals [Figure~\ref{fig1A}(b)]. In the time intervals between revivals, the phase difference, $\Delta \phi \left( t \right)$, changes slightly, but still is not constant in time [Figure~\ref{fig1A}(b)]. The increase in the coupling constant leads to the phase difference begins to change noticeably at all times [Figure~\ref{fig1A}(d) and (f)]. Thus, we can conclude that the phase transition occurring at $\Omega  = {\Omega _{SSE}}$ (see the main text) is not related to the phase locking of the oscillators \cite{ref48}.

\begin{figure}[htbp]
\centering\includegraphics[width=\linewidth]{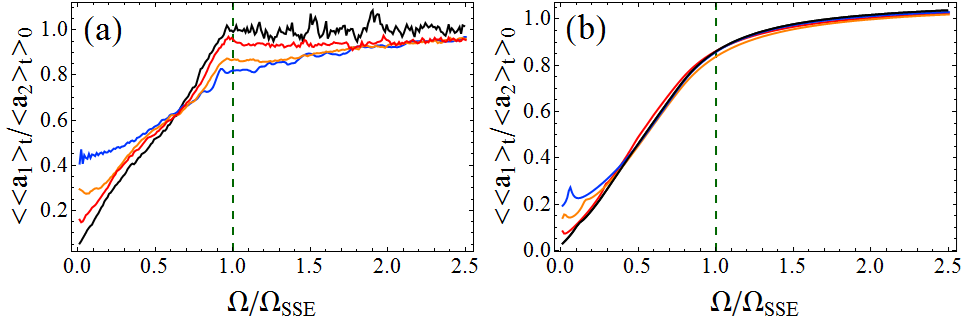}
\caption{The dependence of ${\left\langle {{{\left\langle {\left| {{a_1}} \right|} \right\rangle }_t}/{{\left\langle {\left| {{a_2}} \right|} \right\rangle }_t}} \right\rangle _0}$ on the coupling strength $\Omega $ when the number of oscillators in the set $N = 50$ (the blue line), $N = 100$ (the orange line), $N = 200$ (the red line), $N = 400$ (the black line). The interval of averaging is $\left[ {0,\,\,25\,{T_R}} \right]$ (a) and $\left[ {0,\,\,0.5\,{T_R}} \right]$ (b). The dashed vertical line shows the critical coupling strength ${\Omega _{SSE}} = \gamma /\sqrt 2 $, where  $\gamma  = \pi {g^2}/\delta \omega \approx 0.014\omega_0$. The averaging is carried out over $200$ initial states. When the number of the oscillators ($N$) change, $g$ is scaled as $1/\sqrt N $ and $\delta \omega $ is scaled as $1/N$.}
\label{fig2A}
\end{figure}

The changes in the system dynamics during the increase of $\Omega $ clearly manifest in the dependence of ${\left\langle {{{\left\langle {\left| {{a_1}} \right|} \right\rangle }_t}/{{\left\langle {\left| {{a_2}} \right|} \right\rangle }_t}} \right\rangle _0}$ on the coupling strength $\Omega $, where ${\left\langle {\left| {{a_{1,2}}} \right|} \right\rangle _t} = \frac{1}{{{t_{\max }}}}\int\limits_0^{{t_{\max }}} {\left| {{a_{1,2}}\left( t \right)} \right|dt} $ and ${\left\langle {...} \right\rangle _0}$ denotes averaging over initial conditions. Our calculation shows that when the averaging of ${\left\langle {{{\left\langle {\left| {{a_1}} \right|} \right\rangle }_t}/{{\left\langle {\left| {{a_2}} \right|} \right\rangle }_t}} \right\rangle _0}$ is carried out over time interval containing large number of the collapses and revivals, the dependence of ${\left\langle {{{\left\langle {\left| {{a_1}} \right|} \right\rangle }_t}/{{\left\langle {\left| {{a_2}} \right|} \right\rangle }_t}} \right\rangle _0}$ on the coupling strength demonstrates a clear threshold at $\Omega  = {\Omega _{SSE}}$ [Figure~\ref{fig5} in the main text]. The increase of the number of oscillators in the set ($N \to \infty $ and $\delta \omega /{\omega _0} \to 0$) makes the transition at $\Omega  = {\Omega _{SSE}}$ sharper [Figure~\ref{fig2A}(a) and Figure~\ref{fig5} in the main text]. At the same time, when the averaging of ${\left\langle {{{\left\langle {\left| {{a_1}} \right|} \right\rangle }_t}/{{\left\langle {\left| {{a_2}} \right|} \right\rangle }_t}} \right\rangle _0}$ is carried out over a time interval smaller than the time of first revival, the dependence of ${\left\langle {{{\left\langle {\left| {{a_1}} \right|} \right\rangle }_t}/{{\left\langle {\left| {{a_2}} \right|} \right\rangle }_t}} \right\rangle _0}$ on the coupling strength is smooth [Figure~\ref{fig2A}(b)]. In this case, the increase in $N$ ($N \to \infty $ and $\delta \omega /{\omega _0} \to 0$) does not lead to the appearance of a sharp threshold [Figure~\ref{fig2A}]. This confirms that the predicted transition is due to the Hermitian dynamics of the system at times much longer than the return time.

\section*{Appendix E. Scaling the system parameters}
We study the dependence of ${\left\langle {{{\left\langle {\left| {{a_1}} \right|} \right\rangle }_t}/{{\left\langle {\left| {{a_2}} \right|} \right\rangle }_t}} \right\rangle _0}$ on the number of oscillators in the set, $N$. The frequencies of the oscillators in the set uniformly fill the interval from ${\omega _0} - N\,\delta \omega /2$ to ${\omega _0} + N\,\delta \omega /2$. To keep the frequency interval unchanged, we scale $\delta \omega $ as ${N^{ - 1}}$ ($N\,\delta \omega  = const$, i.e. the density of states in the reservoir increases). The increase in $N$ leads to the enhancement of the energy exchange between the first oscillator and the set of oscillators. Based on the expression for the relaxation rate, $\gamma $, (10), we conclude that such an enhancement is proportional to $1/\delta \omega  \sim N$. To keep the magnitude of the relaxation rate, we scale $g$ (coupling strength between the first oscillator and each of the oscillators in the set) as ${N^{ - 1/2}}$.

\begin{figure}[htbp]
\centering\includegraphics[width=\linewidth]{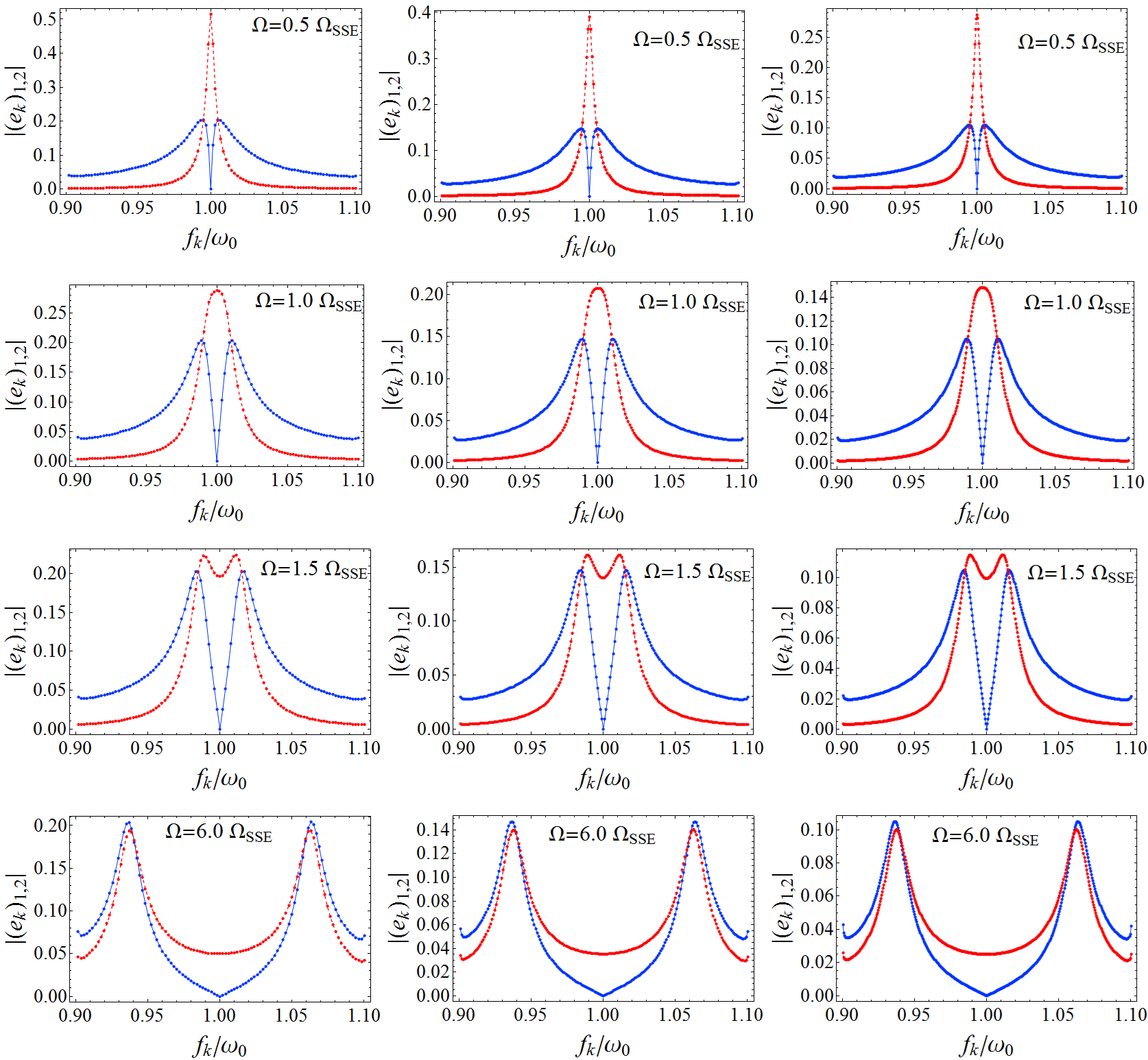}
\caption{The dependence of the amplitudes of first ${\left( {{e_k}} \right)_1}$ (blue points) and second ${\left( {{e_k}} \right)_2}$ (red points) components in the eigenstate ${{\mathbf{e}}_k}$ on its eigenfrequency ${f_k}$ for the different coupling strengths: $\Omega  = 0.5\,{\Omega _{SSE}}$ (first row); $\Omega  = {\Omega _{SSE}}$ (second row); $\Omega  = 1.5\,{\Omega _{SSE}}$ (third row); $\Omega  = 6\,{\Omega _{SSE}}$ (fourth row). Here ${\Omega _{SSE}} = \gamma /\sqrt 2 $ and  $\gamma  = \pi {g^2}/\delta \omega \approx 0.014\omega_0$. The left column is plotted for $100$ oscillators in the set; the center column is plotted for $200$ oscillators in the set; and the right column is plotted for $400$ oscillators in the set. When the number of the oscillators $N$ changes, $g$ is scaled as $1/\sqrt N $ and $\delta \omega $ is scaled as $1/N$.}
\label{fig3A}
\end{figure}

Our numerical simulation shows that the increase in the number of oscillators in the set ($N \to \infty $, $\delta \omega  \sim {N^{ - 1}}$, $g \sim {N^{ - 1/2}}$) makes the transition at $\Omega  = {\Omega _{SSE}}$ sharper [Figure~\ref{fig5} in the main text]. To prove that the phase transition appears due to the change in the eigenstates [Figure~\ref{fig2} in the main text], we calculate the eigenstates for different $N$. Our calculations show that at the parameter scaling described above ($\delta \omega  \sim {N^{ - 1}}$, $g \sim {N^{ - 1/2}}$), an increase in $N$ preserves the behavior of the eigenstates (Figure~\ref{fig3A}). At $\Omega  < {\Omega _{SSE}}$, the value of the second component of the eigenstate with eigenfrequency ${f_k} = {\omega _0}$ is larger than the value of the second component of any other eigenstate [see the first line in Figure~\ref{fig3A}]. In contrast, this eigenstate with ${f_k} = {\omega _0}$ has a zero value of the first component [see the first line in Figure~\ref{fig3A}]. The increase in the coupling strength leads to a splitting of the peak in the dependence of the second components of the eigenstates on their eigenfrequencies. This splitting begins to appear at the critical coupling strength $\Omega  = {\Omega _{SSE}}$ [the second line in Figure~\ref{fig3A}]. At $\Omega  > {\Omega _{SSE}}$, both components of the eigenstates have two peaks [the third and fourth lines in Figure~\ref{fig3A}]. Thus, we conclude that the increase in $N$ does not change the value of $\Omega $ at which the splitting in the eigenstates occurs.
 
\section*{Appendix F. Emergence of symmetry for non-symmetrical eigenstates}
The eigenstates experience the qualitative change at the coupling strength $\Omega  = {\Omega _{SSE}}$. When $\Omega $ increases above ${\Omega _{SSE}}$, the amplitudes of first ${\left( {{e_k}} \right)_1}$ (blue points) and second ${\left( {{e_k}} \right)_2}$ (red points) components in the eigenstate ${{\mathbf{e}}_k}$ become more symmetrical [Figure~\ref{fig3A}]. This symmetrization occurs smoothly. At the same time, the ratio ${\left\langle {{{\left\langle {\left| {{a_1}} \right|} \right\rangle }_t}/{{\left\langle {\left| {{a_2}} \right|} \right\rangle }_t}} \right\rangle _0}$ demonstrates the threshold behavior on the coupling strength $\Omega  = {\Omega _{SSE}}$.

To comprehend the source of the sharp symmetrization of ${\left\langle {{{\left\langle {\left| {{a_1}} \right|} \right\rangle }_t}/{{\left\langle {\left| {{a_2}} \right|} \right\rangle }_t}} \right\rangle _0}$ at $\Omega  = {\Omega _{SSE}}$, we consider in more details the system dynamics. The dynamics of the first and second oscillators can be described by the following formula:
\begin{equation}
{a_{1,2}}(t) = \sum\limits_{k = 1}^{N + 2} {{C_k}{{\left( {{e_k}} \right)}_{1,2}}{e^{i\,{f_k}t}}} 
\label{eq:11A}
\end{equation}
where ${f_k}$ are the eigenfrequencies of the Eqns.~(\ref{eq:3A})-(\ref{eq:5A}), ${\left( {{e_k}} \right)_{1,2}}$ are the first and second components of the eigenstates and ${C_k}$ are the decomposition coefficients of initial conditions by eigenstates, which are given as ${C_k} = a_1^{\left( 0 \right)}\left( {{e_k}} \right)_1^* + a_2^{\left( 0 \right)}\left( {{e_k}} \right)_2^*$, where $a_1^{\left( 0 \right)}$ and $a_2^{\left( 0 \right)}$ are initial amplitudes of the first and second oscillators, respectively. $N$ is a number of the oscillators in the set.

It is important that all ${\left( {{e_k}} \right)_{1,2}}$ can be chosen to be real at the same time. Then the temporal dependences for the amplitudes of first and second oscillators are given as
\begin{equation}
{a_1}(t) = a_1^{\left( 0 \right)}\sum\limits_{k = 1}^{N + 2} {\left( {{e_k}} \right)_1^2{e^{i\,{f_k}t}}}  + a_2^{\left( 0 \right)}\sum\limits_{k = 1}^{N + 2} {{{\left( {{e_k}} \right)}_1}{{\left( {{e_k}} \right)}_2}{e^{i\,{f_k}t}}}
\label{eq:12A}
\end{equation}

\begin{equation}
{a_2}(t) = a_2^{\left( 0 \right)}\sum\limits_{k = 1}^{N + 2} {\left( {{e_k}} \right)_2^2{e^{i\,{f_k}t}}}  + a_1^{\left( 0 \right)}\sum\limits_{k = 1}^{N + 2} {{{\left( {{e_k}} \right)}_1}{{\left( {{e_k}} \right)}_2}{e^{i\,{f_k}t}}}
\label{eq:13A}
\end{equation}
We introduce a notation $\xi \left( t \right) = \sum\limits_{k = 1}^{N + 2} {{{\left( {{e_k}} \right)}_1}{{\left( {{e_k}} \right)}_2}{e^{i\,\left( {{f_k} - {\omega _0}} \right)t}}}$ and find the expressions for the absolute values of the oscillators amplitudes
\begin{equation}
\begin{gathered}
  \left| {{a_1}(t)} \right| = \left( {{{\left| {a_1^{\left( 0 \right)}} \right|}^2}{{\left| {\sum\limits_{k = 1}^{N + 2} {\left( {{e_k}} \right)_1^2{e^{i\,{f_k}t}}} } \right|}^2} + {{\left| {a_2^{\left( 0 \right)}} \right|}^2}{{\left| {\xi \left( t \right)} \right|}^2} + } \right. \hfill \\
  {\left. {2\operatorname{Re} \left( {a_1^{\left( 0 \right)}a_2^{\left( 0 \right)*}\sum\limits_{k = 1}^{N + 2} {\left( {{e_k}} \right)_1^2{e^{i\,{f_k}t}}} {\xi ^*}\left( t \right)} \right)} \right)^{1/2}} \hfill \\ 
\end{gathered}
\label{eq:14A}
\end{equation}

\begin{equation}
\begin{gathered}
  \left| {{a_2}(t)} \right| = \left( {{{\left| {a_2^{\left( 0 \right)}} \right|}^2}{{\left| {\sum\limits_{k = 1}^{N + 2} {\left( {{e_k}} \right)_2^2{e^{i\,{f_k}t}}} } \right|}^2} + {{\left| {a_1^{\left( 0 \right)}} \right|}^2}{{\left| {\xi \left( t \right)} \right|}^2} + } \right. \hfill \\
  {\left. {2\operatorname{Re} \left( {a_2^{\left( 0 \right)}a_1^{\left( 0 \right)*}\sum\limits_{k = 1}^{N + 2} {\left( {{e_k}} \right)_2^2{e^{i\,{f_k}t}}} {\xi ^*}\left( t \right)} \right)} \right)^{1/2}} \hfill \\ 
\end{gathered}
\label{eq:15A}
\end{equation}

\begin{figure}[htbp]
\centering\includegraphics[width=0.7\linewidth]{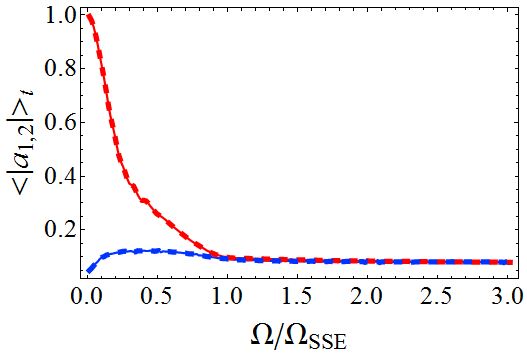}
\caption{The dependences of the amplitudes of the first (blue lines) and second (red lines) oscillators averaged over time on the coupling strength between the oscillators. The solid lines calculated with taking into account the sign-alternating terms in the Eqns.~(\ref{eq:14A}), (\ref{eq:15A}); the dashed lines calculated without taking into account the sign-alternating terms.}
\label{fig4A}
\end{figure}

The last terms in the Eqns.~(\ref{eq:14A}) and (\ref{eq:15A}) are sign-alternating and, when averaging over time, their contributions tend to zero [Figure~\ref{fig4A}]. Therefore, the following equation takes place
\begin{equation}
\frac{{{{\left\langle {\left| {{a_1}} \right|} \right\rangle }_t}}}{{{{\left\langle {\left| {{a_2}} \right|} \right\rangle }_t}}} \approx \frac{{\int\limits_0^T {\sqrt {{{\left| {a_1^{\left( 0 \right)}} \right|}^2}{{\left| {\sum\limits_{k = 1}^{N + 2} {\left( {{e_k}} \right)_1^2{e^{i\,{f_k}t}}} } \right|}^2} + {{\left| {a_2^{\left( 0 \right)}} \right|}^2}{{\left| {\xi \left( t \right)} \right|}^2}} dt} }}{{\int\limits_0^T {\sqrt {{{\left| {a_2^{\left( 0 \right)}} \right|}^2}{{\left| {\sum\limits_{k = 1}^{N + 2} {\left( {{e_k}} \right)_2^2{e^{i\,{f_k}t}}} } \right|}^2} + {{\left| {a_1^{\left( 0 \right)}} \right|}^2}{{\left| {\xi \left( t \right)} \right|}^2}} dt} }}
\label{eq:16A}
\end{equation}

\begin{figure*}[htbp]
\centering\includegraphics[width=\linewidth]{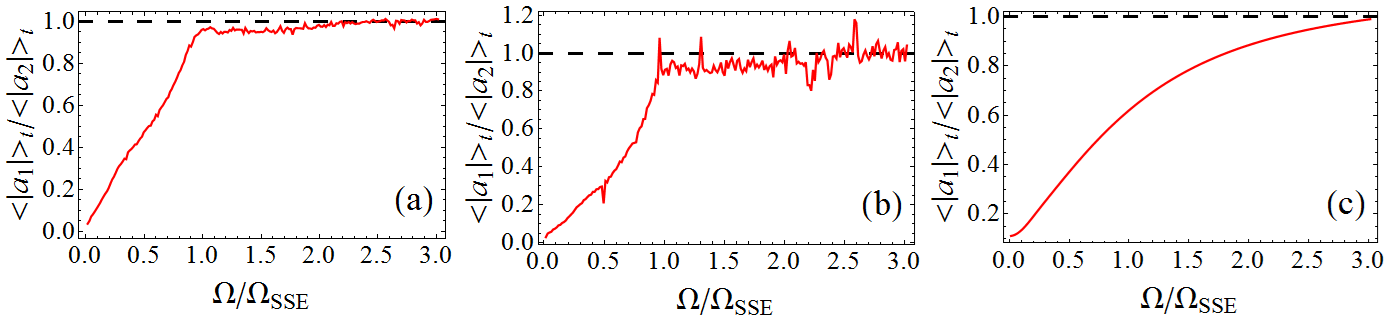}
\caption{(a) The dependence of the ratio ${\left\langle {\left| {{a_1}} \right|} \right\rangle _t}/{\left\langle {\left| {{a_2}} \right|} \right\rangle _t}$ calculated without taking into account the sign-alternating terms in the Eqns.~(\ref{eq:14A}), (\ref{eq:15A}), but with taking into account the term ${\left| {\xi \left( t \right)} \right|^2}$, on the coupling strength between the oscillators [Eq.~(\ref{eq:16A})]. The number of the oscillators in the set is $200$. (b) The dependence of the ratio ${\left\langle {\left| {{a_1}} \right|} \right\rangle _t}/{\left\langle {\left| {{a_2}} \right|} \right\rangle _t}$ calculated without taking into account the sign-alternating terms and without taking into account the term ${\left| {\xi \left( t \right)} \right|^2}$ on the coupling strength, i.e. $\int\limits_0^T {\left| {\sum\limits_{k = 1}^{N + 2} {\left( {{e_k}} \right)_1^2{e^{i\,{f_k}t}}} } \right|dt/\int\limits_0^T {\left| {\sum\limits_{k = 1}^{N + 2} {\left( {{e_k}} \right)_2^2{e^{i\,{f_k}t}}} } \right|dt} }$. (c) The dependence of the ratio $\int\limits_0^T {\sqrt {\sum\limits_{k = 1}^{N + 2} {\left( {{e_k}} \right)_1^4} } dt} /\int\limits_0^T {\sqrt {\sum\limits_{k = 1}^{N + 2} {\left( {{e_k}} \right)_2^4} } dt}$ on the coupling strength.}
\label{fig5A}
\end{figure*}

When averaging over the initial states, the factors ${\left| {a_1^{\left( 0 \right)}} \right|^2}$ and ${\left| {a_2^{\left( 0 \right)}} \right|^2}$ are equal to each other. In this case, the ratio 

\begin{equation}
{\left\langle {\left| {{a_1}} \right|} \right\rangle _t}/{\left\langle {\left| {{a_2}} \right|} \right\rangle _t} \approx \frac{{\int\limits_0^T {\sqrt {{{\left| {\sum\limits_{k = 1}^{N + 2} {\left( {{e_k}} \right)_1^2{e^{i\,{f_k}t}}} } \right|}^2} + {{\left| {\xi \left( t \right)} \right|}^2}} dt} }}{{\int\limits_0^T {\sqrt {{{\left| {\sum\limits_{k = 1}^{N + 2} {\left( {{e_k}} \right)_2^2{e^{i\,{f_k}t}}} } \right|}^2} + {{\left| {\xi \left( t \right)} \right|}^2}} dt} }}
\label{eq:15A1}
\end{equation}
demonstrates a clear transition at $\Omega  = {\Omega _{SSE}}$ [Figure~\ref{fig5A}(a)]. The second terms in the ratio are the same, therefore, this ratio is determined by the relation between ${\left| {\sum\limits_{k = 1}^{N + 2} {\left( {{e_k}} \right)_1^2{e^{i\,{f_k}t}}} } \right|^2}$ and ${\left| {\sum\limits_{k = 1}^{N + 2} {\left( {{e_k}} \right)_2^2{e^{i\,{f_k}t}}} } \right|^2}$.

Our~~~calculations~~~show~~~that~~~the~~~ratio~~(\ref{eq:16A})~~~ without~~the~~second terms,~~i.e., \\$\int\limits_0^T {\left| {\sum\limits_{k = 1}^{N + 2} {\left( {{e_k}} \right)_1^2{e^{i\,{f_k}t}}} } \right|dt/\int\limits_0^T {\left| {\sum\limits_{k = 1}^{N + 2} {\left( {{e_k}} \right)_2^2{e^{i\,{f_k}t}}} } \right|dt} }$, also demonstrates a clear transition at $\Omega  = {\Omega _{SSE}}$ [Figure~\ref{fig5A}(b)]. Thus, we can assume that the transition is due to the changes in the expressions $\int\limits_0^T {\left| {\sum\limits_{k = 1}^{N + 2} {\left( {{e_k}} \right)_{1,2}^2{e^{i\,{f_k}t}}} } \right|dt}$. These expressions can be rewritten as 
\begin{equation}
\begin{gathered}
  \int\limits_0^T {\left| {\sum\limits_{k = 1}^{N + 2} {\left( {{e_k}} \right)_{1,2}^2{e^{i\,{f_k}t}}} } \right|dt}  =  \hfill 
  \int\limits_0^T {{{\left( {\sum\limits_{k = 1}^{N + 2} {\left( {{e_k}} \right)_{1,2}^4 + \sum\limits_{k = 1}^{N + 2} {\sum\limits_{m = 1}^{N + 2} {\left( {{e_k}} \right)_{1,2}^2\left( {{e_m}} \right)_{1,2}^2{e^{i\,\left( {{f_k} - {f_m}} \right)t}}} } } } \right)}^{1/2}}dt}  \hfill \\ 
\end{gathered}
\label{eq:17A}
\end{equation}
where, in the double sum, $k \ne m$.

The first term in the Eq.~(\ref{eq:17A}) does not depend on time, whereas the second term oscillates in time in a complex way. The presence of the oscillating term is necessary for existence of the transition in the considered system. Indeed, our calculation shows that the ratio $\int\limits_0^T {\sqrt {\sum\limits_{k = 1}^{N + 2} {\left( {{e_k}} \right)_1^4} } dt} /\int\limits_0^T {\sqrt {\sum\limits_{k = 1}^{N + 2} {\left( {{e_k}} \right)_2^4} } dt}$ does not demonstrate the threshold behavior and smoothly tends to $1$ when the coupling strength increases [Figure~\ref{fig5A}(c)]. Thus, the interference between the modes plays a key role in the formation of the transition. Note that the total sum $\sum\limits_{k = 1}^{N + 2} {\sum\limits_{m = 1}^{N + 2} {\left( {{e_k}} \right)_{1,2}^2\left( {{e_m}} \right)_{1,2}^2}  = \sum\limits_{k = 1}^{N + 2} {\left( {{e_k}} \right)_{1,2}^2\sum\limits_{m = 1}^{N + 2} {\left( {{e_m}} \right)_{1,2}^2} } }$ is equal to $1$ ($\sum\limits_{m = 1}^{N + 2} {\left( {{e_m}} \right)_{1,2}^2}  = 1$). For this reason, the transition manifests itself only at time much greater than the return time, when the phase difference between the various modes becomes significant [cf. Figure~\ref{fig2A}].
To ascertain the mechanism of the transition appearance, we determine the main contribution in the sums $\left| {\sum\limits_{k = 1}^{N + 2} {\left( {{e_k}} \right)_{1,2}^2{e^{i\,{f_k}t}}} } \right|$. Below the transition point, the second component ${\left( {{e_k}} \right)_2}$ has one peak, whereas, the first component ${\left( {{e_k}} \right)_1}$ has two peaks [the first row in Figure~\ref{fig3A}]. Above the transition point, both components have two peaks. In the case, when the second component has one peak, we can rewrite the expression~(\ref{eq:17A}) in the following form

\begin{equation}
\begin{gathered}
  \left| {\sum\limits_{k = 1}^{N + 2} {\left( {{e_k}} \right)_2^2{e^{i\,{f_k}t}}} } \right| = \left| {{e^{i{\omega _0}t}}\,\sum\limits_{k = 1}^{N + 2} {\left( {{e_k}} \right)_2^2{e^{i\,\left( {{f_k} - {\omega _0}} \right)t}}} } \right| =  \hfill  \left| {\sum\limits_{k = 1}^{N + 2} {\left( {{e_k}} \right)_2^2{e^{i\,\left( {{f_k} - {\omega _0}} \right)t}}} } \right| \hfill \\ 
\end{gathered}
\label{eq:18A}
\end{equation}
where ${\omega _0}$ is the peak frequency (it coincides with the oscillators frequency).

In the case when the component has two peaks, the expression~(\ref{eq:17A}) can be rewritten in the following form
\begin{equation}
\begin{gathered}
  \left| {\sum\limits_{k = 1}^{N + 2} {\left( {{e_k}} \right)_{1,2}^2{e^{i\,{f_k}t}}} } \right| =  \hfill 
  \left| \begin{gathered}
  {e^{i\,\left( {{\omega _0} - \Delta {\omega _{1,2}}} \right)t}}\sum\limits_{k = 1}^{N/2 + 1} {\left( {{e_k}} \right)_{1,2}^2{e^{i\,\left( {{f_k} - {\omega _0} + \Delta {\omega _{1,2}}} \right)t}}}  +  \hfill \\
  {e^{i\,\left( {{\omega _0} + \Delta {\omega _{1,2}}} \right)t}}\sum\limits_{k = N/2 + 2}^{N + 2} {\left( {{e_k}} \right)_{1,2}^2{e^{i\,\left( {{f_k} - {\omega _0} - \Delta {\omega _{1,2}}} \right)t}}}  \hfill \\ 
\end{gathered}  \right| =  \hfill \\
  \left| \begin{gathered}
  {e^{ - i\Delta {\omega _{\max }}t}}\sum\limits_{k = 1}^{N/2 + 1} {\left( {{e_k}} \right)_{1,2}^2{e^{i\,\left( {{f_k} - {\omega _0} + \Delta {\omega _{1,2}}} \right)t}}}  +  \hfill \\
  {e^{i\,\Delta {\omega _{\max }}t}}\sum\limits_{k = N/2 + 2}^{N + 2} {\left( {{e_k}} \right)_{1,2}^2{e^{i\,\left( {{f_k} - {\omega _0} - \Delta {\omega _{1,2}}} \right)t}}}  \hfill \\ 
\end{gathered}  \right| \hfill \\ 
\end{gathered}
\label{eq:19A}
\end{equation}

On the right part of the Eq.~(\ref{eq:19A}) the first sum includes only the eigenstates with the eigenfrequencies ${f_k} < {\omega _0}$ and the second sum includes only the eigenstates with the eigenfrequencies ${f_k} > {\omega _0}$. ${\omega _0} \pm \Delta {\omega _{1,2}}$ are the frequencies of peaks for the corresponding components [Figure~\ref{fig3A}].
Taking into account the dependences of ${\left( {{e_k}} \right)_{1,2}}$ on ${f_k}$, we conclude that 
\begin{equation}
\begin{gathered}
  \sum\limits_{k = 1}^{N/2 + 1} {\left( {{e_k}} \right)_{1,2}^2{e^{i\,\left( {{f_k} - {\omega _0} + \Delta {\omega _{\max }}} \right)t}}}  =  \hfill 
  {\left( { - 1} \right)^{1,2}}{\left( {\sum\limits_{k = N/2 + 2}^{N + 2} {\left( {{e_k}} \right)_{1,2}^2{e^{i\,\left( {{f_k} - {\omega _0} - \Delta {\omega _{\max }}} \right)t}}} } \right)^*} \hfill \\ 
\end{gathered}
\label{eq:20A}
\end{equation}
Here $\left(  - 1  \right) ^ {1,2}$ denotes that for the first component the sign is a minus and for the second component the sign is a plus.

Then, we make the following transformation
\begin{equation}
\begin{gathered}
  \left| {\sum\limits_{k = 1}^{N + 2} {\left( {{e_k}} \right)_{1,2}^2{e^{i\,{f_k}t}}} } \right| =  \hfill 
  \left| \begin{gathered}
  {e^{ - i\Delta {\omega _{\max }}t}}\sum\limits_{k = 1}^{N/2 + 1} {\left( {{e_k}} \right)_{1,2}^2{e^{i\,\left( {{f_k} - {\omega _0} + \Delta {\omega _{1,2}}} \right)t}}}  +  \hfill \\
  {\left( { - 1} \right)^{1,2}}{e^{i\,\Delta {\omega _{1,2}}t}}{\left( {\sum\limits_{k = 1}^{N/2 + 1} {\left( {{e_k}} \right)_{1,2}^2{e^{i\,\left( {{f_k} - {\omega _0} + \Delta {\omega _{1,2}}} \right)t}}} } \right)^*} \hfill \\ 
\end{gathered}  \right| =  \hfill \\
  2\left( {{{\left| {\sum\limits_{k = 1}^{N/2 + 1} {\left( {{e_k}} \right)_{1,2}^2{e^{i\,\left( {{f_k} - {\omega _0} + \Delta {\omega _{1,2}}} \right)t}}} } \right|}^2} + } \right.{\left( { \pm 1} \right)^{1,2}} \times  \hfill \\
  {\left. {\operatorname{Re} \left( {{e^{ - 2i\,\Delta {\omega _{1,2}}t}}{{\left( {\sum\limits_{k = 1}^{N/2 + 1} {\left( {{e_k}} \right)_{1,2}^2{e^{i\,\left( {{f_k} - {\omega _0} + \Delta {\omega _{1,2}}} \right)t}}} } \right)}^2}} \right)} \right)^{1/2}} \hfill \\ 
\end{gathered}
\label{eq:21A}
\end{equation}

The second term in the Eq.~(\ref{eq:21A}) is sign-alternating and, when averaging over time, their contribution tends to zero. Thus, we obtain the following expressions for the component, which has two peaks
\begin{equation}
\left| {\sum\limits_{k = 1}^{N + 2} {\left( {{e_k}} \right)_{1,2}^2{e^{i\,{f_k}t}}} } \right| \approx \sqrt 2 \left| {\sum\limits_{k = 1}^{N/2 + 1} {\left( {{e_k}} \right)_{1,2}^2{e^{i\,\left( {{f_k} - {\omega _0} + \Delta {\omega _{1,2}}} \right)t}}} } \right|
\label{eq:22A}
\end{equation}

When $t >  > {T_R}$, due to the eigenfrequencies ${f_k}$ are not equidistant precisely, the various eigenstates are added to each other with arbitrary phases. The maximal contributions in the sums~(\ref{eq:18A}) and (\ref{eq:22A}) are given by the eigenstates with the greatest values of $\left( {{e_k}} \right)_{1,2}^2$, i.e., the eigenstates, the eigenfrequencies of which lie at the peaks.

It is important that in the case when the eigenstates are added to each other with arbitrary phases ($t >  > {T_R}$), the sums~(\ref{eq:18A}) and (\ref{eq:22A}) do not depend on the peak frequencies and depend only on the peak form. In the simplest consideration, the peaks are characterized only by the height, ${P_{1,2}} = \max \left( {{e_k}} \right)_{1,2}^2$, and widths, ${\Gamma _{1,2}}$. Within this approximation, the sums~(\ref{eq:18A}) and (\ref{eq:22A}) are only functions of ${P_{1,2}}$ and ${\Gamma _{1,2}}$. It is clear that the sums~(\ref{eq:18A}) and (\ref{eq:22A}) are linearly proportional to ${P_{1,2}} = \max \left( {{e_k}} \right)_{1,2}^2$. At the same time, since the various eigenstates are added to each other with arbitrary phases, then the sums~(\ref{eq:18A}) and (\ref{eq:22A}) are proportional to the square root of number of the eigenstates within the peak \cite{petrov1976}. This number is proportional to the peak widths and, therefore, the sums~(\ref{eq:18A}) and (\ref{eq:22A}) are proportional to $\sqrt {{\Gamma _{1,2}}}$.

Below the transition point, the second component has one peak and the first component has two peaks, and, therefore,
\begin{equation}
\left| {\sum\limits_{k = 1}^{N + 2} {\left( {{e_k}} \right)_2^2{e^{i\,{f_k}t}}} } \right| \approx \eta \left( {{P_2},{\Gamma _2}} \right) \sim \sqrt {{\Gamma _2}} \,{P_2}
\label{eq:23A}
\end{equation}

\begin{equation}
\left| {\sum\limits_{k = 1}^{N + 2} {\left( {{e_k}} \right)_1^2{e^{i\,{f_k}t}}} } \right| \approx \sqrt 2 \,\eta \left( {{P_1},{\Gamma _1}} \right) \sim \sqrt {2\,{\Gamma _1}} \,{P_1}
\label{eq:24A}
\end{equation}

Above the transition point
\begin{equation}
\left| {\sum\limits_{k = 1}^{N + 2} {\left( {{e_k}} \right)_2^2{e^{i\,{f_k}t}}} } \right| \approx \sqrt 2 \,\eta \left( {{P_2},{\Gamma _2}} \right) \sim \sqrt {2\,{\Gamma _2}} \,{P_2}
\label{eq:25A}
\end{equation}

\begin{equation}
\left| {\sum\limits_{k = 1}^{N + 2} {\left( {{e_k}} \right)_1^2{e^{i\,{f_k}t}}} } \right| \approx \sqrt 2 \,\eta \left( {{P_1},{\Gamma _1}} \right) \sim \sqrt {2\,{\Gamma _1}} \,{P_1}
\label{eq:26A}
\end{equation}

It can be seen that when the component has two peaks, the expression~(\ref{eq:24A}) is proportional to the square root of total width of the two peaks. Therefore, to estimate the sum~(\ref{eq:22A}) at the transition point, at which the one peak splits into the two peaks, we use the expression~(\ref{eq:23A}) considering that ${\Gamma _2}$ is a total width of the split peak and begin to use the expression~(\ref{eq:25A}) , when the height of minimum between the two peaks becomes less than the half of absolute height.

Using the Eqns.~(\ref{eq:23A})--(\ref{eq:26A}), we calculate the dependences of the sums~(\ref{eq:18A}), (\ref{eq:22A}) on the coupling strength. It is important that the expressions~(\ref{eq:23A})--(\ref{eq:26A}) do not depend on time and, therefore, the averaging over time does not change the answer.

\begin{figure*}[htbp]
\centering\includegraphics[width=\linewidth]{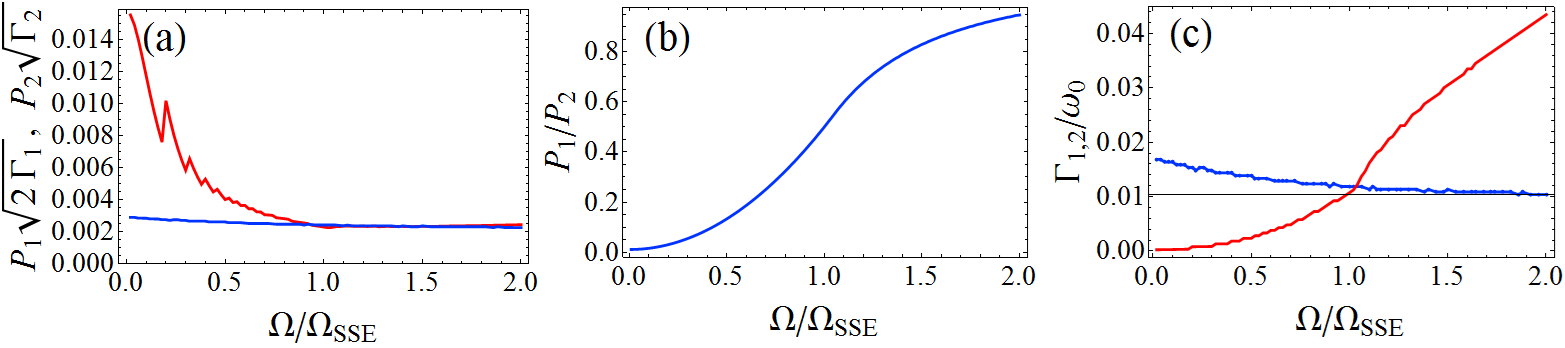}
\caption{(a) The dependences of $\left| {\sum\limits_{k = 1}^N {\left( {{e_k}} \right)_{1,2}^2{e^{i\,{f_k}t}}} } \right|$ (the blue line) and $\left| {\sum\limits_{k = 1}^N {\left( {{e_k}} \right)_2^2{e^{i\,{f_k}t}}} } \right|$ (the red line) calculated by the Eqns.~(\ref{eq:23A})--(\ref{eq:26A}) on the coupling strength between the oscillators. (b) The dependence of the ratio of peak heights, ${P_1}/{P_2} = \max \left( {{e_k}} \right)_1^2/\max \left( {{e_k}} \right)_2^2$, on the coupling strength. (c) The dependence of the peak widths, ${\Gamma _{1,2}}$, on the coupling strength. ${\Gamma _1}$ is shown by the blue line, and ${\Gamma _2}$ is shown by the red line.}
\label{fig6A}
\end{figure*}

Our calculations show that when the coupling strength $\Omega  > {\Omega _{SSE}}$, $\left| {\sum\limits_{k = 1}^N {\left( {{e_k}} \right)_1^2{e^{i\,{f_k}t}}} } \right|$ (the blue line) and $\left| {\sum\limits_{k = 1}^N {\left( {{e_k}} \right)_2^2{e^{i\,{f_k}t}}} } \right|$ (the red line) calculated by the Eqns.~(\ref{eq:23A})--(\ref{eq:26A}) are almost exactly equal to each other [Figure~\ref{fig6A}(a)]. At the same time, ${P_1} \ne {P_2}$ and ${\Gamma _1} \ne {\Gamma _2}$ almost for all values of the coupling strength [Figure~\ref{fig6A}(b), (c)], and the first and second components of the eigenstates differ from each other. Thus, the sums $\left| {\sum\limits_{k = 1}^N {\left( {{e_k}} \right)_{1,2}^2{e^{i\,{f_k}t}}} } \right|$ for the first and second components becomes the same, whereas, the first and second components are not the same.

When the eigenstates are added to each other with arbitrary phases, the sums~(\ref{eq:18A}) and (\ref{eq:22A}) can be treated as random variables. The ratio of the variance of such a random variable to the averaged value decreases as $1/\sqrt N $. For this reason, the increase in the number of oscillators in the set, $N$, leads to a decrease in the relative fluctuations of each of the sums and the ratio of these sums becomes more exactly equal to $1$. Moreover, it can be seen from Figure~\ref{fig3A} that an increase in the number of oscillators, $N$, leads only to an increase in the density of the eigenstates. Therefore, in the limit of a large number of oscillators, $N$, a set of eigenstates can be described by a continuous function to which the set converges. In this limit, the approximations~(\ref{eq:23A})--(\ref{eq:26A}) become more accurate, and the ratio of the sums~(\ref{eq:18A}) and (\ref{eq:22A}) tends to $1$.
 
The value of ${\left\langle {\left| {{a_1}} \right|} \right\rangle _t}/{\left\langle {\left| {{a_2}} \right|} \right\rangle _t}$ is determined by the ratio of $\left| {\sum\limits_{k = 1}^N {\left( {{e_k}} \right)_1^2{e^{i\,{f_k}t}}} } \right|/\left| {\sum\limits_{k = 1}^N {\left( {{e_k}} \right)_2^2{e^{i\,{f_k}t}}} } \right|$ [Eq.~(\ref{eq:16A})]. Therefore, when this ratio tends to $1$, the ratio of ${\left\langle {{{\left\langle {\left| {{a_1}} \right|} \right\rangle }_t}/{{\left\langle {\left| {{a_2}} \right|} \right\rangle }_t}} \right\rangle _0}$ also comes to $1$ [Figure~\ref{fig2A}(a)].

Thus, the dependence of ${\left\langle {{{\left\langle {\left| {{a_1}} \right|} \right\rangle }_t}/{{\left\langle {\left| {{a_2}} \right|} \right\rangle }_t}} \right\rangle _0}$ on the coupling strength demonstrates a clear threshold at $\Omega  = {\Omega _{SSE}}$ [see Figure~\ref{fig2A}(a)] and the system dynamics becomes symmetrical with respect to this characteristic, although the first and second components of the eigenstates are not symmetrical [Figure~\ref{fig3A}]. This is due to the fact that when $t >  > {T_R}$, the averages ${\left\langle {\left| {{a_1}} \right|} \right\rangle _t}$ and ${\left\langle {\left| {{a_2}} \right|} \right\rangle _t}$ are determined by the sums of the eigenstates' components, which are added to each other with arbitrary phases. These sums are proportional to $\sqrt {{\Gamma _{1,2}}} \,{P_{1,2}}$, which are determined only by the eigenstates and become equal to each other above the transition point.

Note that when $t < {T_R}$, the phases of the eigenstates components in the sums~(\ref{eq:18A}) and (\ref{eq:22A}) are not random, therefore, the sums are not proportional to $\sqrt {{\Gamma _{1,2}}} \,{P_{1,2}}$ and the dependence of ${\left\langle {{{\left\langle {\left| {{a_1}} \right|} \right\rangle }_t}/{{\left\langle {\left| {{a_2}} \right|} \right\rangle }_t}} \right\rangle _0}$ on the coupling is smooth [Figure~\ref{fig2A}(b)]. Moreover, when the averaging time is much less than the return time ${T_R}$, the averaging over time does not lead to zeroing of the sign-alternating terms in the Eqns.~(\ref{eq:14A}), (\ref{eq:15A}). This is also a reason that the dependence of ${\left\langle {{{\left\langle {\left| {{a_1}} \right|} \right\rangle }_t}/{{\left\langle {\left| {{a_2}} \right|} \right\rangle }_t}} \right\rangle _0}$ on the coupling strength is smooth [Figure~\ref{fig2A}(b)].

\end{document}